\documentclass[twocolumn,english,aps,prd,nofootinbib]{revtex4}
\usepackage[latin9]{inputenc}
\usepackage{amsmath}
\usepackage{amssymb}
\usepackage{graphicx}
\usepackage{esint}

\makeatletter

\DeclareFontEncoding{LGR}{}{}
\DeclareRobustCommand{\greektext}{%
  \fontencoding{LGR}\selectfont\def\encodingdefault{LGR}}
\DeclareRobustCommand{\textgreek}[1]{\leavevmode{\greektext #1}}
\ProvideTextCommand{\~}{LGR}[1]{\char126#1}

\newcommand{\lyxmathsym}[1]{\ifmmode\begingroup\def\b@ld{bold}
  \text{\ifx\math@version\b@ld\bfseries\fi#1}\endgroup\else#1\fi}

\DeclareTextSymbolDefault{\textquotedbl}{T1}

\@ifundefined{textcolor}{}
{%
 \definecolor{BLACK}{gray}{0}
 \definecolor{WHITE}{gray}{1}
 \definecolor{RED}{rgb}{1,0,0}
 \definecolor{GREEN}{rgb}{0,1,0}
 \definecolor{BLUE}{rgb}{0,0,1}
 \definecolor{CYAN}{cmyk}{1,0,0,0}
 \definecolor{MAGENTA}{cmyk}{0,1,0,0}
 \definecolor{YELLOW}{cmyk}{0,0,1,0}
}

\@ifundefined{definecolor}
 {\usepackage{color}}{}

\usepackage{enumerate}

\makeatother

\usepackage{babel}

\makeatother

\usepackage{babel}
\begin{document}
\title{Analyzing Pantheon SNeIa data in the context of Barrow's variable
speed of light}
\author{Hoang Ky Nguyen$\,$}
\email[\ \ ]{hoang.nguyen@ubbcluj.ro}
\affiliation{Department of Physics, Babe\c{s}-Bolyai University, Cluj-Napoca 400084, Romania}

\date{April 4, 2024}

\begin{abstract}
\noindent We analyze the Combined Pantheon Sample of Type Ia supernovae
while allowing the velocity of light to vary as a function of the
scale factor $c\propto a^{-\zeta}$, as initiated by Barrow [Phys.$\,$Rev.$\,$D \textbf{59}, 043515 (1999)].
The variation in the velocity of light creates an effect akin to the
refraction phenomenon which occurs for a wave traveling in a medium
with varying speed of wave. We elucidate the role of the local scale
of gravitationally-bound regions \emph{in assisting the refraction
effect to manifest}. The refraction effect alters the redshift formulae
(Lema\^itre, distance-vs-$z$, luminosity distance-vs-$z$) and warrants
a new analysis of the Pantheon dataset. Upon a reformulation of the
distance-redshift relations, we achieve a high-quality fit of the
Pantheon dataset to the variable light speed approach; the fit is
as robust as that obtained in the standard $\Lambda CDM$ model. We
find that the Pantheon dataset is consistent with the variable light
speed of the functional form: $a\propto t^{\mu}$ and $c\propto a^{1-1/\mu}$
with (i) the cosmic age {\normalsize{}$t_{0}\approx13.9$} Gy as a
free parameter, while $\mu$ is unspecified; and (ii) a monotonic
variation in the local scale for gravitationally-bound objects (applicable
to the emission sources and the Solar System-based apparatus/observer).
Due to the agent in (ii), the high-$z$ portion of the Pantheon dataset
would produce an ``effective'' $H_{0}$ estimate which is $10$
percent lower than the $H_{0}$ estimate obtained from the low-$z$
portion of the dataset. We offer an alternative interpretation of
the accelerating expansion by way of variable speed of light, and
as a by-product of the agent uncovered in (ii), a tentative suggestion
toward ``resolving'' the ongoing tension in the Hubble constant
estimates.
\end{abstract}
\maketitle

\section{\label{sec:motivation}motivation}

\noindent The possibility of variation in the velocity of light was
first advocated by Einstein in 1911 \citep{Einstein1911} in his search
for a formulation of General Relativity (GR). As he emphasized in
\citep{Einstein1912-1,Einstein1912-2}, his consideration of variable
velocity of light is not in contradiction with the principle of the
constancy of the velocity of light. This is because the latter, or
equivalently the Michelson-Morley experimental result and the Lorentz
invariance, is meant to be valid only \emph{locally}. At a given point
on the manifold, the set of tangent frames satisfy the Lorentz invariance
with a common value of $c$. However, Einstein recognized that the
value of $c$ in principle may vary on the manifold. In the language
of the geometry, whereas the speed of light is an \emph{invariant}
(meaning that $c$ is unaffected upon a general coordinate transformation),
it can be \emph{position-dependent}, viz. $c(x^{\mu})$. The speed
of light in principle can be a scalar field, rather than a universal
constant. In Ref. \citep{Einstein1911}, Einstein explicitly allowed
the gravitational field to influence the value of $c$. \footnote{Note that the speed of light appears in two places: (1) in the underlying
theory as an \emph{invariant} and (2) in the metric which depends
on the choice of ruler and clock. The speed of light that participates
in the underlying theory is what Einstein intended and is the focus
of our Report. We thank Viktor Toth for clarifying the distinct roles
of $c$ in the two places, (1) vs (2).}

The modern form of variable speed of light (VSL hereafter) was revived
in the work of Moffat in 1992 \citep{Moffat} and independently by
Albrecht and Magueijo in 1998 \citep{Magueijo1} in the context of
early-time cosmology. Their proposals aimed to resolve the horizon
puzzle while avoiding the need for cosmic inflation. Several scholars
explored different aspects of the VSL, most notably being \citep{Barrow3,Barrow1,Barrow2,Magueijo2,Magueijo3,Magueijo4}.

The application of VSL in late-time cosmology, mostly to analyzing
the Type Ia supernovae (SNeIa) data, has been more limited, apparently
without clear successes \citep{Zhang,Qi,Ravanpak,Salzano}. Upon reviewing
these works, we conclude that they continued to rely on the classic
Lema\^itre redshift formula, viz. $1+z=a^{-1}$. We believe that
this is an oversight, however. There are certain types of VSL in which
the classic Lema\^itre redshift formula is no longer applicable,
warranting a revision. One such type of VSL exists in the form of
the velocity of light being dependent on the scale factor, e.g. $c\propto a^{-\zeta}$,
a form first employed by Barrow \citep{Barrow3}. As will be shown
in Section \ref{subsec:Modifying-Lemaitre-formula} in this Report,
as a traveling lightwave makes a transit between a gravitationally-bound
region which resists cosmic expansion and an outer space region that
is subject to cosmic expansion, \emph{proper care is needed to handle
the alteration of the wavelength}. \footnote{Previous VSL works overlooked an additional alteration in the wavelength
which takes place when the lightwave transits from the outer space
region into the (gravitationally-bound) Milky Way to reach the Earth-based
observer. See Section \ref{subsec:Modifying-Lemaitre-formula} and
Appendix \ref{sec:Equivalent-derivation} for our detailed rectification
of the oversight.} We find that transits of this type do introduce fundamental modifications
to Lema\^itre's redshift formula\emph{. }In particular, Barrow's
VSL form $c\propto a^{-\zeta}$ would result in a \emph{modified}
Lema\^itre formula: $1+z=a^{-(1+\zeta)}$. We believe that the lack
of progress in applying VSL to late-time cosmology has been due to
the aforementioned oversight in missing out the VSL component (in
Barrow's case, the exponent $\zeta$) in the redshift relations. It
is the purpose of our Report to properly sort out this intricacy. 

In this Report, we apply Barrow's VSL form $c\propto a^{-\zeta}$
into analyzing the SNeIa data and interpreting the accelerating expansion.
As a lightwave travels from a distant emitter toward the Earth-based
observer, if the velocity of light varies as a function the scale
factor (with the latter meaning the global scale factor of the cosmos
\emph{and} the local scale factor of gravitationally-bound regions),
the lightwave would undergo a refraction effect akin to the phenomenon
which takes place as a physical wave travels in an inhomogeneous medium
with varying speed of wave. That is how the exponent $\zeta$ finds
its way into the modified Lema\^itre redshift formula, $1+z=a^{-(1+\zeta)}$,
alluded above. To our knowledge, this is the first time the refraction
effect is explicitly considered for late-time cosmology.

The trove of data in the Combined Pantheon Sample \citep{Scolnic,Pantheon-data}
enables our current study. The question we aim to address is whether
the refraction effect induced by way of the VSL alone can account
for the Pantheon dataset and provide a complete explanation of the
accelerating expansion observed in SNeIa \citep{Riess1,Perlmutter}.
This pursuit is legitimate: although GR has been verified in the scale
of the solar system or that of binary stars, the Friedmann equations
involve an extrapolation of GR onto the cosmic scales. This extrapolation
constitutes a major assumption in cosmology \citep{Bull}. As observational
data of SNeIa did not fit with the original Friedmann framework, among
other reasons, standard cosmology introduced new components such as
the $\Lambda$ term to account for the accelerating expansion. Our
analysis presented in this Report does not resort to GR or the Friedmann
equations. It is based on a quite generic and parsimonious setup without
relying on an underlying theory of gravitation.

Our Report is structured as follows. In Section \ref{sec:Modified-RW}
we introduce a VSL modification in the Robertson-Walker metric. In
Section \ref{sec:Modified-redshift-formulae} we derive the modified
redshift relations (Lema\^itre, distance-vs-$z$, luminosity distance-vs-$z$)
by enabling variation in the speed of light and variation in the local
scale of the gravitationally-bound regions; \emph{we also elucidate
the role of the latter in assisting the effect of VSL on the redshift
to manifest}. In Section \ref{sec:Fitting-of-modified-formulae} we
conduct an analysis of the Combined Pantheon Sample using our modified
luminosity distance-vs-redshift formula derived in the preceding Section.
In Section \ref{sec:Interpretations-of-results} we discuss the implications
of our Pantheon analysis; in particular, we provide (i) a new interpretation
of the accelerating expansion; and (ii) the potential explanation
to the Hubble constant tension by way of the variation in the local
scale of gravitationally-bound regions. Appendix \ref{sec:Equivalent-derivation}
provides an alternative route to the modified Lema\^itre redshift
formula derived in Section \ref{sec:Modified-redshift-formulae}.

\section{\label{sec:Modified-RW}The modified Robertson-Walker metric}

\noindent The Robertson-Walker (RW hereafter) metric starts with the
assumption of homogeneity and isotropy of space. It also assumes the
spatial component of the metric to be time-dependent. All of the time
dependence is in the function $a(t)$ known as the cosmic scale factor.
The RW metric is the only one that is spatially homogeneous and isotropic.
This is a geometrical result and is not tied to the equations of the
gravitational field.

The RW metric has been determined to be:
\begin{align}
ds^{2} & =c^{2}dt^{2}-a^{2}(t)\left[\frac{dr^{2}}{1-kr^{2}}+r^{2}d\Omega^{2}\right]\label{eq:RW}\\
d\Omega^{2} & =d\theta^{2}+\sin^{2}\theta\,d\phi^{2}
\end{align}
where the global cosmic scale factor $a(t)$ is a function of the
cosmic time $t$ only (with $a(t_{0})=1$ at our current time $t_{0}$),
and $k$ the curvature determining the shape of the universe (open/flat/closed
for $k>0,k=0,k<0$ respectively). The quantity $c$ in the RW metric
is taken to be the velocity of light $c$ in Einstein's underlying
theory; the latter $c$ is an invariant (i.e., unaffected under a
general coordinate transformation) but in principle can vary on the
manifold. In the language of geometry, $c$ can be a scalar field,
rather than a universal constant.

$ $

When allowing for the variation in the velocity of light on the manifold,
we retain the homogeneity and isotropy of space. In Barrow's VSL form
\citep{Barrow3}, the velocity of light is a function of the scale
factor, viz.
\begin{equation}
c=c_{0}\,a^{-\zeta}\label{eq:Barrow-func-1}
\end{equation}
The RW metric remains applicable with only a minor modification being
in the dependence of $c$ on the scale factor. We thus arrive at the
\emph{modified} RW metric:
\begin{align}
ds^{2} & =c^{2}(a)\,dt^{2}-a^{2}(t)\left[\frac{dr^{2}}{1-kr^{2}}+r^{2}d\Omega^{2}\right]\label{eq:modified-RW}\\
 & =\frac{c_{0}^{2}}{a^{2\zeta}(t)}\,dt^{2}-a^{2}(t)\left[\frac{dr^{2}}{1-kr^{2}}+r^{2}d\Omega^{2}\right]\label{eq:modified-RW-zeta}
\end{align}
in which $c_{0}$ is the speed of light measured in the outer space
region (which is subject to cosmic expansion) at our current time.

\section{\label{sec:Modified-redshift-formulae}Modifying redshift formulae
to enable variation in the velocity of light}

\subsection{\label{subsec:Role-grav-bound}Role of gravitationally-bound regions
in detecting the redshift}

\noindent The RW metric deals with the global cosmic scale factor.
However, in order for an observer to detect the redshift, the region
containing the observer must not be directly subject to cosmic expansion.
\footnote{\noindent If the Solar System had expanded together with the cosmos,
the observer's apparatus would also have expanded in sync with the
wavelength of the light ray emitted from a distant supernova and could
not have detected any redshift. } Thanks to the attraction nature of gravity, regions that are populated
with matter are expected to be able to resist cosmic expansion and
retain their own local scale over the course of time since their formation
post the recombination event. In order to understand the impact of
variable velocity of light on the redshift, it is necessary for us
to explicitly consider the local scale of gravitationally-bound regions.
The reason is as follows.

A lightwave emitted from a supernova in a distant (gravitationally-bound)
galaxy first has to transit into the outer space region which is not
gravitationally-bound. It will then traverse the null geodesic of
the RW metric and expands together with the cosmic factor $a(t)$.
Finally, the lightwave must transit into the (gravitationally-bound)
Milky Way to reach the astronomer's apparatus. Whereas the middle
stage is well understood in standard cosmology, the first and the
last stages have received little attention. It turns out that the
first and the last stages are crucial in the context of VSL. This
Section prepares the notations and concepts needed for our detailed
examination in Section \ref{subsec:Refraction-effect}.

\noindent 
\begin{figure*}[!t]
\noindent \begin{centering}
\includegraphics[scale=0.85]{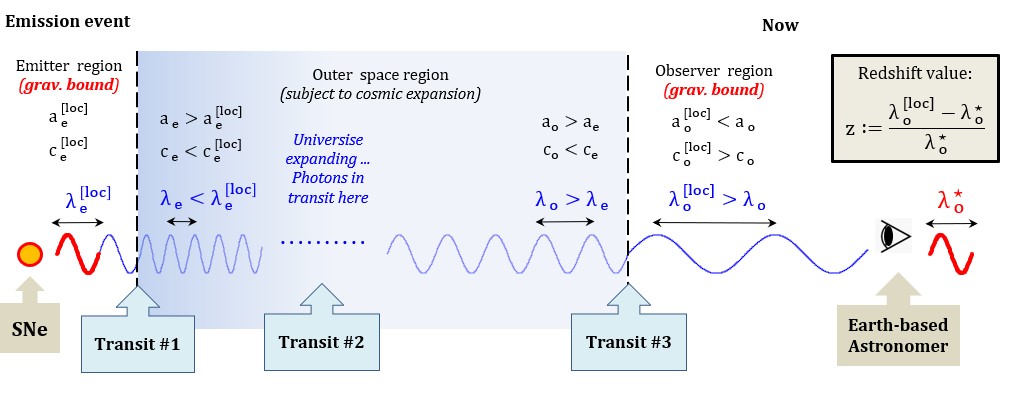}
\par\end{centering}
\caption{Demonstrating the case of $c\propto a^{-1/2}$. A lightwave from a
supernova emission (shown on the far left) will have made 3 transits
to reach the Earth-based astronomer (shown on the far right). In Transit
\#1, the lightwave exits the (gravitationally-bound) emission galaxy
to enter the outer space region that surrounds the emission galaxy;
the wavecrest gets compressed as light slows down. During Transit
\#2, the lightwave travels in the outer space which undergoes a cosmic
expansion; as a result, the lightwave expands. In Transit \#3, the
lightwave enters the (gravitationally-bound) Milky Way; the wavecrest
expands further as light speeds up. The Earth-based astronomer measures
$\lambda_{o}^{[loc]}$ and compares it with the yardstick $\lambda_{o}^{*}$
in order to produce the redshift value (shown in the upper right corner
box).}

\label{fig:wavetrain-full}
\end{figure*}

Figure \ref{fig:wavetrain-full} on Page \pageref{fig:wavetrain-full}
gives us a quick glimpse of the issue. On its way from a distant supernova
to the Earth-based astronomer, the light ray must undergo 3 transits:
\begin{itemize}
\item Transit $\#$1: from the \emph{gravitationally-bound} galaxy (which
contains the supernova) to the outer space region that encloses the
emission galaxy.
\item Transit $\#$2: from the outer space region that encloses the emission
galaxy to the outer space region that encloses the Milky Way. During
Transit \#2, the wavecrest follows the null geodesic of the RW metric
and expands as a result of the cosmic expansion.
\item Transit $\#$3: from the outer space region that encloses the Milky
Way into the \emph{gravitationally-bound} Milky Way to finally reach
the Earth-based observer.
\end{itemize}
\noindent As shown in Figure \ref{fig:wavetrain-full}, let us denote:
\begin{itemize}
\item $a_{e}^{[loc]}$, the local scale of the emission galaxy;
\item $a_{o}^{[loc]}$, the local scale of the Milky Way;
\item $a_{e}$, the global cosmic scale of the outer space region that encloses
the emission galaxy;
\item $a_{o}$, the global cosmic scale of the outer space region that encloses
the Milky Way;
\end{itemize}
in which the subscript $\lyxmathsym{\textquotedblleft}e\lyxmathsym{\textquotedblright}$
and $\lyxmathsym{\textquotedblleft}o\lyxmathsym{\textquotedblright}$
stand for ``emission'' and ``observation'' respectively, and the
superscript $[loc]$ means ``local''. Note that, in principle, $a_{e}^{[loc]}$
and $a_{o}^{[loc]}$ may be \emph{different}. 

$ $

\noindent Likewise, let us denote:
\begin{itemize}
\item $\lambda{}_{e}^{[loc]}$, the wavelength of the photon emitted by
the supernova;
\item $\lambda{}_{o}^{[loc]}$, the wavelength of the photon detected by
the observer;
\item $\lambda_{e}$, the wavelength of the photon in the outer space region
that encloses the emission galaxy;
\item $\lambda_{o}$, the wavelength of the photon in the outer space region
that encloses the Milky Way.
\end{itemize}
In addition, $\lambda_{o}^{*}$ is the wavelength of the photon \emph{if
the source were to emit the photon at the observer's location}. It
serves as a yardstick against which the Earth-based astronomer compares
her detected wavelength $\lambda_{o}^{[loc]}$ and produce the value
of the redshift. In standard cosmology: 
\begin{equation}
\lambda_{o}^{*}=\lambda_{e}^{[loc]}
\end{equation}
In the context of VSL, it is possible and necessary to set $\lambda_{o}^{*}$
and $\lambda_{e}^{[loc]}$ apart. We let $\lambda_{o}^{*}$ be related
to $\lambda_{e}^{[loc]}$ via the local scale of the emission galaxy
and that of the Milky Way per the following relation:
\begin{equation}
\frac{\lambda_{o}^{*}}{\lambda_{e}^{[loc]}}=\frac{a_{o}^{[loc]}}{a_{e}^{[loc]}}\label{eq:definition-of-yardstick}
\end{equation}
As will become clear in Section \ref{subsec:Role-of-local-scale},
by making the local scale of the emitter, $a_{e}^{[loc]}$, be dependent
on its redshift $z$ and exploiting Relation \eqref{eq:definition-of-yardstick},
we would obtain a potential explanation to the Hubble constant tension
(i.e, the discrepancy in the estimates of $H_{0}$ from late-time
objects \citep{tension1,tension2,tension3} versus from the Cosmic
Microwave Background \citep{tension4,tension10,tension11,tension12,tension13,tension9,tension8,tension7,tension6,tension5}).
This is a by-product benefit of our analysis.

Finally, the Earth-based astronomer measures $\lambda_{o}^{[loc]}$
then compares it with $\lambda_{o}^{*}$ to obtain the value of the
redshift:
\begin{equation}
z:=\frac{\lambda_{o}^{[loc]}-\lambda_{o}^{*}}{\lambda_{o}^{*}}\label{eq:definition-of-z}
\end{equation}

\subsection{\label{subsec:Refraction-effect}A refraction effect due to variable
speed of light}

\noindent 
\begin{figure}
\begin{centering}
\includegraphics[scale=0.7]{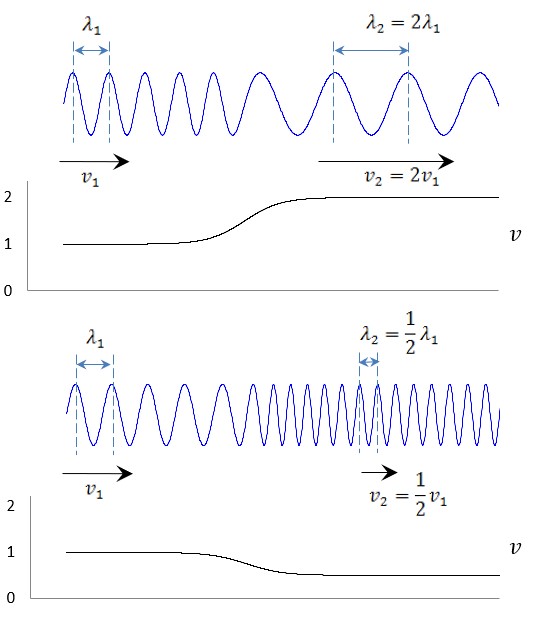}
\par\end{centering}
\caption{Change in wavelength as a wave travels in medium with varying speed
of wave. Upper panel: as its speed doubles, so does its wavelength.
Lower panel: wavelength halves as its speed halves, so does its wavelength.
In either case, wavelength and velocity are proportional: $\lambda_{2}/v_{2}=\lambda_{1}/v_{1}$.}

\label{fig:wavetrains}
\end{figure}

\noindent In the VSL context depicted in Fig. \ref{fig:wavetrain-full},
we further denote:
\begin{itemize}
\item $c_{e}^{[loc]}$, the speed of light inside the emission galaxy (which
has the local factor $a_{e}^{[loc]}$);
\item $c_{o}^{[loc]}$, the speed of light inside the Milky Way (which has
the local factor $a_{o}^{[loc]}$). Note that $c_{o}^{[loc]}=c_{SS}=300,000\,km/sec$
is the speed of light measured in the Solar System where the astronomer
lives. The ``SS'' in $c_{SS}$ stands for ``Solar System''.
\item $c_{e}$, the speed of light in the outer space region (which has
the global factor $a_{e}$) that encloses the emission galaxy;
\item $c_{o}$, the speed of light in the outer space region (which has
the global factor $a_{o}$) that encloses the Milky Way.
\end{itemize}
\noindent Let us start with a well-understood phenomenon: the behavior
of a light ray in a medium with varying refractive index. It is well
established that the wavelength of the light ray at a given location
is proportional to the velocity of light at that location:

\noindent 
\begin{equation}
\lambda\propto v\label{eq:lambda-v-speed}
\end{equation}
Figure \ref{fig:wavetrains} illustrates the change in wavelength
as a wave travels at varying velocity. In the upper panel, as the
velocity increases, the front end of the wavecrest will rush forward
leaving its back end behind thus stretching out the wavecrest. In
the lower panel, the reverse situation occurs: as the velocity decreases,
the front end of the wavecrest will slow down while its back end continues
its course thus compressing the wavecrest. In either situation, the
wavelength and the velocity of wave are directly proportional:
\begin{equation}
\frac{\lambda_{2}}{\lambda_{1}}=\frac{v_{2}}{v_{1}}
\end{equation}

$ $

\noindent For our problem at hand, the refraction phenomenon also
applies to lightwaves which travel from a distant supernova toward
the Earth-based astronomer. Figure \ref{fig:wavetrain-full} illustrates
the essence of the refraction effect. It shows how the wavelength
observed on Earth acquires an extra factor as compared with a classic
case in which the speed of light is universal. For concreteness, Figure
\ref{fig:wavetrain-full} demonstrates the example $c\propto a^{-1/2}$,
namely, the speed of light decreases in reverse of the scale factor.

In Figure \ref{fig:wavetrain-full}, as a light ray travels from distant
SNe (far left) to Earth-based astronomer (far right), it passes through
4 regions, each one having its own scale and velocity of light. The
4 regions are:\renewcommand{\labelenumi}{\arabic{enumi})}
\begin{enumerate}
\item The (gravitationally-bound) emission galaxy: scale $a_{e}^{[loc]}$
and velocity of light $c_{e}^{[loc]}$. The photon emitted at this
event has wavelength $\lambda_{e}^{[loc]}$.
\item The outer space region enclosing the emission galaxy: scale $a_{e}$
and velocity of light $c_{e}$. The outer space region is subject
to cosmic expansion. Since the outer space region might already have
expanded before the SNe emits the light ray, in general:
\begin{equation}
a_{e}\geqslant a_{e}^{[loc]}
\end{equation}
\item The outer space region enclosing the Milky Way: scale $a_{o}$ and
velocity of light $c_{o}$. As a result of cosmic expansion, the following
inequality holds:
\begin{equation}
a_{o}>a_{e}
\end{equation}
\item Inside the (gravitationally-bound) Solar System where the Earth-based
observer resides: scale $a_{o}^{[loc]}$ and velocity of light $c_{o}^{[loc]}$.
Note that it is $c_{o}^{[loc]}=c_{SS}=300,000\,km/sec$, with ``SS''
in $c_{SS}$ short-handing for ``Solar System''. As such, the following
inequality holds:
\begin{equation}
a_{o}^{[loc]}<a_{o}
\end{equation}
\end{enumerate}
\noindent \renewcommand{\labelenumi}{\arabic{enumi}.}As a concrete
example, Figure \ref{fig:wavetrain-full} sets $a_{e}=3\,a_{e}^{[loc]};\ a_{o}=2\,a_{e};\ a_{o}^{[loc]}=\frac{1}{6}\,a_{o}$.
In this example, $a_{o}^{[loc]}=a_{e}^{[loc]}$.

$ $

$ $

\noindent The lightwave from a supernova must make 3 transits before
reaching the Earth-based astronomer.

\noindent \textbf{$ $}

\noindent \textbf{Transit $\#$1:}

\noindent $ $

\noindent The lightwave exits the emission galaxy (which is gravitationally-bound)
to enter the outer space region (which is subject to cosmic expansion)
that surrounds the emission galaxy. During Transit \#1, the wavecrest
gets compressed as light slows down, viz. $c_{e}<c_{e}^{[loc]}$ due
to $a_{e}>a_{e}^{[loc]}$. In Fig. \ref{fig:wavetrain-full}, $a_{e}=3\,a{}_{e}^{[loc]}$
leading to $c_{e}=\frac{1}{\sqrt{3}}\,c_{e}^{[loc]}$ per $c\propto a^{-1/2}$,
and $\lambda_{e}=\frac{1}{\sqrt{3}}\,\lambda{}_{e}^{[loc]}$ as a
result of the refraction effect per $\lambda\propto c$.

\noindent \textbf{$ $}

\noindent \textbf{Transit $\#$2:}

\noindent $ $

\noindent The lightwave traverse the null geodesic of the RW metric,
from the outer space region that encloses the emission galaxy to the
outskirt of the Milky Way. During Transit \#2, the wavecrest gets
stretched out as a result of the cosmic expansion: $a_{o}>a_{e}$.
In Fig. \ref{fig:wavetrain-full}, $a_{o}=2\,a_{e}$ leading to $c_{o}=\frac{1}{\sqrt{2}}\,c_{e}$
per $c\propto a^{-1/2}$, and $\lambda_{o}=2\,\lambda_{e}$ resulting
from the cosmic expansion.

\noindent \textbf{$ $}

\noindent \textbf{Transit $\#$3:}

\noindent $ $

\noindent From the outskirt of the Milky Way, the lightwave enters
the Milky Way (which is gravitationally-bound) and finally reaches
the Earth-based observer. During Transit \#3, the wavecrest gets stretched
out further as the light speed increases: $c_{o}^{[loc]}>c_{o}$ due
to $a_{o}^{[loc]}<a_{o}$. In Fig. \ref{fig:wavetrain-full}, $a_{o}^{[loc]}=\frac{1}{6}\,a_{o}$
leading to $c_{o}^{[loc]}=\sqrt{6}\,c_{o}$ per $c\propto a^{-1/2}$,
and $\lambda_{o}^{[loc]}=\sqrt{6}\,\lambda_{o}$ as a result of the
refraction effect per $\lambda\propto c$.

\noindent \textbf{$ $}

\noindent \textbf{The net effect:}

\noindent $ $

\noindent The Earth-based astronomer compares her observation $\lambda_{o}^{[loc]}$
with $\lambda_{o}^{*}$ both of which are \emph{directly measurable}
by the astronomer. She would find that $\lambda_{o}^{[loc]}=\sqrt{6}\,\lambda_{o}=\sqrt{6}\times2\,\lambda_{e}=\sqrt{6}\times2\times\frac{1}{\sqrt{3}}\,\lambda_{e}^{[loc]}=2^{3/2}\,\lambda_{e}^{[loc]}$.
Note that in this example: $\lambda_{o}^{*}=\lambda_{e}^{[loc]}$
because $a_{o}^{[loc]}=a_{e}^{[loc]}$. The astronomer would thus
find that $\lambda_{o}^{[loc]}=2^{3/2}\,\lambda_{o}^{*}$.

$ $

Note that in standard cosmology, the astronomer would only find that
$\lambda_{o}^{[loc]}=2\,\lambda_{o}^{*}$ which is a direct result
of the cosmic expansion, viz. $a_{o}=2\,a_{e}$, taking place during
Transit \#2. In standard cosmology, as light speed is non-varying,
Transit \#1 and Transit \#3 do not affect the wavelength of the light
ray.

$ $

The Earth-based astronomer measures $\lambda_{o}^{[loc]}$ then compares
it with $\lambda_{o}^{*}$ to obtain the value of the redshift:
\begin{equation}
z:=\frac{\lambda_{o}^{[loc]}-\lambda_{o}^{*}}{\lambda_{o}^{*}}
\end{equation}
In the example depicted in Fig. \ref{fig:wavetrain-full}, the astronomer
thus obtains a redshift value of $z_{VSL}=2^{3/2}-1\approx1.82$.
In the absence of VSL, she would obtain a redshift value of $z_{standard}=2-1=1$
only.

$ $

Generally speaking, for a decreasing function of $c$ w.r.t. the scale
factor $a$, the VSL yields a larger redshift than what is produced
in standard cosmology. In the context of VSL, the change in the redshift
therefore warrants a revision in the distance-vs-redshift relations,
the topics we shall expound in the rest of this Section.

\subsection{\label{subsec:Modifying-Lemaitre-formula}Modifying Lema\^itre's
formula}

\noindent What is most interesting in the demonstration in Figure
\ref{fig:wavetrain-full} is that the stretching out of the wavecrest
during Transit \#3 does \emph{not} cancel out the compression of the
wavecrest during Transit \#1. The refraction occurring at Transit
\#1 and the ``reverse'' refraction taking place at Transit \#3 do
not net each other out. For a decreasing function of $c$, the net
effect increases the value of $z$ and results in a new formula for
the redshift. Below is our derivation.

$ $

\noindent Due to the refraction effect at Transit \#1:
\begin{equation}
\lambda_{e}=\lambda_{e}^{[loc]}\frac{c_{e}}{c_{e}^{[loc]}}\label{eq:Transit1}
\end{equation}
Due to the cosmic expansion during Transit \#2:
\begin{equation}
\lambda_{o}=\lambda_{e}\frac{a_{o}}{a_{e}}\label{eq:Transit2}
\end{equation}
Due to the \textquotedbl reverse'' refraction effect at Transit
\#3:
\begin{equation}
\lambda_{o}^{[loc]}=\lambda_{o}\frac{c_{o}^{[loc]}}{c_{o}}\label{eq:Transit3}
\end{equation}
Combining (\ref{eq:Transit1}-\ref{eq:Transit3}), we have:
\begin{equation}
\lambda_{o}^{[loc]}=\lambda_{e}^{[loc]}\frac{c_{e}}{c_{e}^{[loc]}}.\frac{a_{o}}{a_{e}}.\frac{c_{o}^{[loc]}}{c_{o}}
\end{equation}
Further combining it with the yardstick, viz. \eqref{eq:definition-of-yardstick},
and the definition of the redshift in \eqref{eq:definition-of-z},
we have:
\begin{align}
1+z & :=\frac{\lambda_{o}^{[loc]}}{\lambda_{o}^{*}}\\
 & =\frac{\lambda_{o}^{[loc]}}{\lambda_{e}^{[loc]}}.\frac{\lambda_{e}^{[loc]}}{\lambda_{o}^{\star}}\\
 & =\left(\frac{c_{e}}{c_{e}^{[loc]}}.\frac{a_{o}}{a_{e}}.\frac{c_{o}^{[loc]}}{c_{o}}\right)\frac{a_{e}^{[loc]}}{a_{o}^{[loc]}}\\
 & =\left(\frac{c_{e}}{c_{o}}.\frac{a_{o}}{a_{e}}\right)\left(\frac{c_{o}^{[loc]}}{c_{e}^{[loc]}}.\frac{a_{e}^{[loc]}}{a_{o}^{[loc]}}\right)\label{eq:tmp-10}
\end{align}
In \citep{Barrow3}, Barrow considered the VSL functional form:
\begin{equation}
c\propto a^{-\zeta}\label{eq:Barrow-func}
\end{equation}
which does not have a a preferred scale. What is new in our approach
is that we enable Barrow's functional form for both types of scale
factor, whether it is global and local. \footnote{The scale factor can be made tightly related to the (Ricci) scalar
curvature. In such a scenario, the speed of light is directly determined
by the Ricci scalar, both of which are invariants. This is the topics
of our follow-up theoretical report \citep{Nguyen-theoretical}. This
result translates into our current report that the applicability of
\eqref{eq:Barrow-func} is non-discriminatory, regardless of whether
the scale factor is global or local.}

$ $

\noindent Applying \eqref{eq:Barrow-func} for the global scale, we
obtain:
\begin{equation}
\frac{c_{e}}{c_{o}}=\left(\frac{a_{e}}{a_{o}}\right)^{-\zeta}\label{eq:tmp11}
\end{equation}
Likewise, applying \eqref{eq:Barrow-func} to the local scale, we
obtain:
\begin{equation}
\frac{c_{o}^{[loc]}}{c_{e}^{[loc]}}=\left(\frac{a_{o}^{[loc]}}{a_{e}^{[loc]}}\right)^{-\zeta}\label{eq:tmp12}
\end{equation}
Combining \eqref{eq:tmp-10}, \eqref{eq:tmp11}, and \eqref{eq:tmp12},
we arrive at the \emph{modified} Lema\^itre redshift formula:
\begin{equation}
1+z=\left(\frac{a_{e}}{a_{o}}\right)^{-(1+\zeta)}\,\left(\frac{a_{e}^{[loc]}}{a_{o}^{[loc]}}\right)^{1+\zeta}\label{eq:modified-Lemaitre-1}
\end{equation}
In \eqref{eq:modified-Lemaitre-1}, $a_{e}^{[loc]}$ is a function
of the redshift $z$ of the emission galaxy. But it may also contains
the idiosyncratic characters of the emission galaxy. Eq. \eqref{eq:modified-Lemaitre-1}
is thus understood in the ``average'' sense in which the idiosyncratic
is being ``averaged'' out. Let us define a new quantity:
\begin{equation}
F(z):=\frac{a_{e}^{[loc]}}{a_{o}^{[loc]}}\label{eq:definition-of-F}
\end{equation}
which is a function of $z$. The \emph{modified} Lema\^itre redshift
formula is (setting $a_{o}=1$):
\begin{equation}
1+z=a_{e}^{-(1+\zeta)}\,F^{1+\zeta}(z)\label{eq:modified-Lemaitre-2}
\end{equation}
The quantity $F(z)$ should be a monotonic and slowly-decreasing function
w.r.t. $z$, satisfying $F(z=0)=1$.

$ $

In the absence of the variation in the local scale, viz. $F(z)\equiv1\ \forall z$,
the modified Lema\^itre redshift formula is simplified to:
\begin{equation}
1+z=a_{e}^{-(1+\zeta)}\label{eq:tmp-8}
\end{equation}
which is fundamentally \emph{different} from the classic Lema\^itre
redshift formula $1+z=a_{e}^{-1}$.

$ $

The VSL exponent $\zeta$ thus has come to the fore: via Eq. \eqref{eq:tmp-8}
it enforces major revisions in the distance-redshift relations and
warrants a new analysis for the Type Ia SNe data. \footnote{Let us verify the demonstration in Figure \ref{fig:wavetrain-full}
in the light of Formula \eqref{eq:tmp-8}. In Figure \ref{fig:wavetrain-full},
$c$ varies as $a^{-1/2}$ (thus $\zeta=1/2$) and the universe has
expanded by a factor of 2, viz. $a_{o}=2\,a_{e}.$ Equivalently, $a\equiv a_{e}=1/2$
(recalling that $a_{o}=1$). Formula \eqref{eq:tmp-8} produces a
redshift $z=(1/2)^{-3/2}-1\approx1.82$ in agreement with the value
of $z_{VSL}$ reported in the last paragraph of Section \ref{subsec:Refraction-effect}.} 

$ $

Let us conclude this subsection by commenting on the oversight in
previous analyses to date \citep{Zhang,Qi,Ravanpak}. In these works,
it was correctly observed that Eq. \eqref{eq:Transit2} is valid regardless
of whether or not the speed of light varies during Transit \#2. However,
this fact alone is not sufficient for the conclusion (used in \citep{Zhang,Qi,Ravanpak})
that the classic Lema\^itre redshift formula, viz. $1+z=a^{-1}$,
should remain valid for VSL. The reason is that $\lambda_{o}$ is
\emph{not} what the Earth-based astronomer observes. To reach the
astronomer, the lightwave needs to enter the Milky Way which has a
scale \emph{smaller} than the current global cosmic scale because
the Milky Way has resisted the cosmic expansion. Since $c\propto a^{-\zeta}$,
the velocity of light $c_{o}^{[loc]}$ inside the Milky Way is \emph{different}
from the velocity of light $c_{o}$ in the outer space region that
encloses the Milky Way. The lightwave thus gets refracted during its
entry to the Milky Way with its wavelength getting altered to $\lambda_{o}^{[loc]}=\lambda_{o}\,c_{o}^{[loc]}/c_{o}$
per Eq. \eqref{eq:Transit3}. It is the wavelength $\lambda_{o}^{[loc]}$
that gets measured in the astronomer's apparatus.

$ $

The novelty of our work is in our exposition of the VSL exponent $\zeta$
in the modified Lema\^itre redshift formula, \eqref{eq:modified-Lemaitre-2}
or \eqref{eq:tmp-8}.

\subsection{\label{subsec:Modifying-d-z}Modifying the distance-redshift formula}

\noindent To ease the notation, from here on, we shall drop the subscript
\textquotedbl$e$\textquotedbl{} and simply write $a$ for $a_{e}$,
$a^{[loc]}$ for $a_{e}^{[loc]}$, $c$ for $c_{e}$, $c^{[loc]}$
for $c_{e}^{[loc]}$. Likewise, we shall write $a_{0}$ for $a_{o}$
(which is conveniently set to $1$), $a_{SS}$ for $a_{o}^{[loc]}$,
$c_{0}$ for $c_{o}$, $c_{SS}$ for $c_{o}^{[loc]}$ (note: $c_{SS}=300,000\,km/sec$)
with ``\emph{SS}'' standing for ``Solar System''.

$ $

In this subsection, we shall derive the distance-redshift formula
applicable for VSL. To proceed, we shall adopt an evolution of the
cosmic scale factor in the functional form:
\begin{equation}
a(t)=\left(\frac{t}{t_{0}}\right)^{\mu}\label{eq:evolution}
\end{equation}
with $a(t_{0})$ being set equal $1$. This functional form has no
preferred scale. It covers the critical mode of expansion for the
flat CDM universe when $\mu=2/3$. Taking derivative of \eqref{eq:evolution},
we obtain the Hubble ``constant'':
\begin{equation}
H(t):=\frac{\dot{a}}{a}=\frac{\mu}{t}
\end{equation}
from which, the cosmic age is:
\begin{equation}
t_{0}=\frac{\mu}{H_{0}}\label{eq:age-formula}
\end{equation}
whereas the total ``proper length'' is:
\begin{align}
s_{0} & =\int_{0}^{t_{0}}dt\,c(a)=c_{0}\int_{0}^{t_{0}}dt\,a^{-\zeta}\nonumber \\
 & =c_{0}\int_{0}^{t_{0}}dt\,\left(\frac{t}{t_{0}}\right)^{-\zeta\,\mu}=\frac{1}{1-\zeta\,\mu}\,c_{0}t_{0}\label{eq:proper-length}
\end{align}
Some useful rearrangement:
\begin{equation}
\frac{\dot{a}}{a}=H=\frac{\mu}{t}=\frac{\mu}{t_{0}}\frac{t_{0}}{t}=H_{0}\frac{1}{a^{1/\mu}}
\end{equation}
From \eqref{eq:Barrow-func-1} and \eqref{eq:modified-RW} with $k=0$,
the coordinate distance in the almost flat space is:
\begin{equation}
r\approx\int_{t_{e}}^{t_{o}}\frac{c(a)\,dt}{a(t)}=c_{0}\int_{t_{e}}^{t_{o}}\frac{dt}{a^{1+\zeta}(t)}\label{eq:coordinate-distance}
\end{equation}
From the modified Lema\^itre redshift formula \eqref{eq:modified-Lemaitre-2},
we get:
\begin{equation}
dz=-(1+\zeta)\frac{\dot{a}}{a^{2+\zeta}}F^{1+\zeta}(z)\,dt
\end{equation}
or
\begin{equation}
\frac{dt}{a^{1+\zeta}}=-\frac{1}{1+\zeta}\frac{dz}{\dot{a}/a}F^{-(1+\zeta)}(z)
\end{equation}
in which we assume that $\dot{a}^{[loc]}/a^{[loc]}\ll\dot{a}/a$.
Eq. \eqref{eq:coordinate-distance} becomes:
\begin{align}
r & =\frac{c_{0}}{1+\zeta}\int_{0}^{z}\frac{dz'}{(\dot{a}/a)(z')}F^{-(1+\zeta)}(z')\nonumber \\
 & =\frac{c_{0}}{(1+\zeta)H_{0}}\int_{0}^{z}dz'a^{1/\mu}F^{-(1+\zeta)}(z')\label{eq:tmp-1}
\end{align}
Combined with the modified Lema\^itre redshift formula \eqref{eq:modified-Lemaitre-2},
the integrand in \eqref{eq:tmp-1} is transformed to:
\begin{equation}
a^{1/\mu}F^{-(1+\zeta)}=\left[a^{1+\zeta}F^{-(1+\zeta)}\right]^{\frac{1}{\mu(1+\zeta)}}F^{\frac{1}{\mu}-(1+\zeta)}\label{eq:integrand}
\end{equation}
Let us define a new parameter $\epsilon$ from $\zeta$ and $\mu$:
\footnote{The total ``proper length'' in \eqref{eq:proper-length} can be
re-expressed as:
\begin{equation}
s_{0}=\frac{1}{\mu+\epsilon/(1+\epsilon)}\,c_{0}t_{0}
\end{equation}
If $\epsilon=0$, from \eqref{eq:age-formula}, the total ``proper
length'' further becomes $\frac{1}{\mu}c_{0}t_{0}=\frac{c_{0}}{H_{0}}\equiv l_{Hubble}$
which is the Hubble length, defined using today's speed of light $c_{0}$.}
\begin{equation}
1+\epsilon:=\frac{1}{\mu(1+\zeta)}\label{eq:definition-of-epsilon}
\end{equation}
or, equivalently:
\begin{equation}
\frac{1}{\mu}-(1+\zeta)=\epsilon\,(1+\zeta)
\end{equation}
upon which Eq. \eqref{eq:integrand} can be conveniently recast, with
the aid of \eqref{eq:modified-Lemaitre-2}, as:
\begin{align}
a^{1/\mu}F^{-(1+\zeta)} & =\left[a^{1+\zeta}F^{-(1+\zeta)}\right]^{1+\epsilon}F^{\epsilon\,(1+\zeta)}\nonumber \\
 & =\frac{1}{(1+z)^{1+\epsilon}}F^{\epsilon\,(1+\zeta)}\label{eq:tmp-2}
\end{align}
By virtue of Eqs. \eqref{eq:tmp-1} and \eqref{eq:tmp-2}, we then
obtain the \emph{modified} distance-redshift relation:
\begin{equation}
\frac{r}{c_{0}}=\frac{1}{(1+\zeta)\,H_{0}}\,\int_{0}^{z}dz'\frac{F^{\epsilon\,(1+\zeta)}(z')}{(1+z')^{1+\epsilon}}
\end{equation}
or, upon applying the age formula \eqref{eq:age-formula}:
\begin{equation}
\frac{r}{c_{0}}=t_{0}\,(1+\epsilon)\,\int_{0}^{z}dz'\frac{F^{\frac{\epsilon}{\mu(1+\epsilon)}}(z')}{(1+z')^{1+\epsilon}}\label{eq:modified-r-vs-z}
\end{equation}
In the limit $\epsilon\rightarrow0$, $F(z)$ disappears from the
integrand in Eq. \eqref{eq:modified-r-vs-z}, and the coordinate distance
is vastly simplified to:
\begin{equation}
\frac{r}{c_{0}}=t_{0}\,\ln(1+z)
\end{equation}
In the limit $\epsilon\rightarrow0$, the two exponents $\zeta$ and
$\mu$ are directly related:
\begin{equation}
1+\zeta=\frac{1}{\mu}
\end{equation}

\noindent For the sake of comparison, we cite the $\Lambda CDM$ distance-redshift
relation (with $\Omega_{curv}=0$):
\begin{equation}
\frac{r}{c}=\frac{1}{H_{0}}\int_{0}^{z}\frac{dz'}{\sqrt{\Omega_{M}(1+z')^{3}+\Omega_{\Lambda}}}
\end{equation}
and that of the flat CDM model ($\Omega_{M}=1,\,\Omega_{\Lambda}=0,\,\Omega_{curve}=0$):
\begin{equation}
\frac{r}{c}=\frac{2}{H_{0}}\left(1-\frac{1}{\sqrt{1+z}}\right)
\end{equation}

\subsection{\label{subsec:Modifying-dL-z}Modifying the luminosity distance-redshift
formula}

\noindent As in standard cosmology, the luminosity distance $d_{L}$
is defined via the absolute luminosity $L$ and the apparent luminosity
$J$:
\begin{equation}
d_{L}^{2}=\frac{L}{4\pi J}\label{eq:definition-of-dL}
\end{equation}
On the other hand, the absolute luminosity $L$ and the apparent luminosity
$J$ are related:
\begin{equation}
4\pi r^{2}J=L\frac{\lambda_{o}^{[loc]}}{\lambda_{e}^{[loc]}}.\frac{\lambda_{o}^{[loc]}}{\lambda_{e}^{[loc]}}\label{eq:J-vs-L}
\end{equation}
In the RHS of \eqref{eq:J-vs-L}, the first term $\lambda_{o}^{[loc]}/\lambda_{e}^{[loc]}$
represents the \textquotedblleft loss\textquotedblright{} in the energy
of the red-shifted photon known as the \textquotedblleft Doppler theft\textquotedblright .
The second (identical) term $\lambda_{o}^{[loc]}/\lambda_{e}^{[loc]}$
is due to the dilution factor in the photon density as the same number
of photons get distributed in a prolonged wavecrest in the radial
direction (i.e., the light ray). The $4\pi r^{2}$ in the LHS of \eqref{eq:J-vs-L}
is the spherical dilution in flat space. From \eqref{eq:definition-of-dL}
and \eqref{eq:J-vs-L}, we have:
\begin{equation}
d_{L}=r\,\frac{\lambda_{o}^{[loc]}}{\lambda_{e}^{[loc]}}
\end{equation}
In terms of the redshift \eqref{eq:definition-of-z} and the yardstick
\eqref{eq:definition-of-yardstick}, the luminosity distance is:
\begin{equation}
d_{L}=r\,\frac{\lambda_{o}^{[loc]}}{\lambda_{o}^{\star}}.\frac{\lambda_{o}^{\star}}{\lambda_{e}^{[loc]}}=r\,(1+z)\,\frac{a_{o}^{[loc]}}{a_{e}^{[loc]}}
\end{equation}
or, by including \eqref{eq:definition-of-F}:
\begin{equation}
d_{L}=r\,(1+z)\,\frac{1}{F(z)}\label{eq:tmp-7}
\end{equation}

\noindent Due to the refraction effect at Transit \#3, the apparent
luminosity distance observed by the Earth-based astronomer $d_{L}^{[loc]}$
differs from $d_{L}$ by the factor $c_{SS}/c_{0}$, viz.

\begin{align}
\frac{d_{L}^{[loc]}}{c_{SS}} & =\frac{d_{L}}{c_{0}}
\end{align}
Finally, combining with \eqref{eq:modified-r-vs-z}, we obtain the
\emph{modified }luminosity distance-redshift relation:
\begin{equation}
\frac{d_{L}^{[loc]}}{c_{SS}}=\frac{t_{0}(1+\epsilon)}{F(z)}\,(1+z)\int_{0}^{z}dz'\frac{F^{\frac{\epsilon}{\mu(1+\epsilon)}}(z')}{(1+z')^{1+\epsilon}}\label{eq:modified-dL-vs-z}
\end{equation}
This is the central formula of our approach to assess Type Ia supernovae
data, to be conducted in Section \ref{subsec:General-fit}. Formula
\eqref{eq:modified-dL-vs-z} contains 3 parameters: $t_{0}$, $\epsilon$
and $\mu$, and involves a function $F(z)$ which captures the dependence
of the local scale of gravitationally-bound regions on the redshift.
The function $F$ cannot be determined from the Friedmann equations
which deal only with the global cosmic scale factor. It would require
separate modeling.

$ $

One potential candidate for the variation of the local scale as a
function of redshift is the following function:
\begin{equation}
F(z):=F_{\infty}+(1-F_{\infty})\frac{2}{1+e^{\kappa z}}\label{eq:F-func}
\end{equation}
This parsimonious choice ensures that $F(z)$ is a monotonic slowly-varying
function, satisfying $F(z=0)=1$. The function in \eqref{eq:F-func}
has 2 parameters: $F_{\infty}$ which is the saturation value of $F$
at high-$z$ and $\kappa$ which specifies the crossover point between
low-$z$ and high-$z$.

For completeness, we cite the standard luminosity distance-redshift
relation (with $\Omega_{curv}=0$):
\begin{equation}
\frac{d_{L}}{c}=\frac{1}{H_{0}}(1+z)\int_{0}^{z}\frac{dz'}{\sqrt{\Omega_{M}(1+z')^{3}+\Omega_{\Lambda}}}\label{eq:LCDM-formula}
\end{equation}
and that in the flat CDM model ($\Omega_{M}=1,\,\Omega_{\Lambda}=0,\,\Omega_{curve}=0$):
\begin{equation}
\frac{d_{L}}{c}=\frac{2}{H_{0}}(1+z)\left(1-\frac{1}{\sqrt{1+z}}\right)\label{eq:flatCDM-formula}
\end{equation}
Note that the flat CDM model \eqref{eq:flatCDM-formula} is a special
case of our VSL formula \eqref{eq:modified-dL-vs-z} when $\epsilon=1/2,\ \mu=2/3$
(hence $\zeta=0$ per \eqref{eq:definition-of-epsilon}) and $F(z)\equiv1\ \forall z$,
and by virtue of $t_{0}=2/(3H_{0})$.

\section{\label{sec:Fitting-of-modified-formulae}Fitting of the modified
redshift formulae to the Pantheon dataset}

\subsection{\label{subsec:General-fit}General assessment}

\noindent We are now well-equipped to assess the observational data
of Type Ia supernovae \citep{Perlmutter,Riess1,Scolnic,Pantheon-data}.
Our numerical tool and the output of our analysis presented in this
Section are available upon request.
\begin{figure}[!t]
\begin{centering}
\includegraphics[scale=0.52]{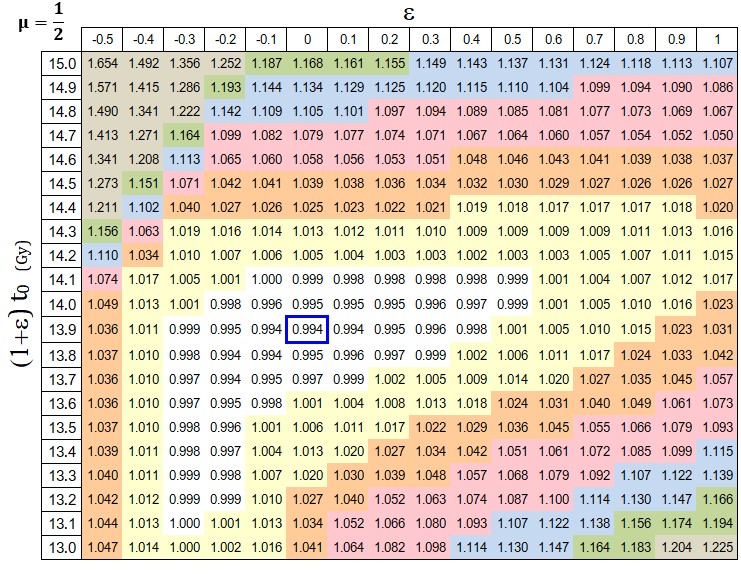}
\par\end{centering}
\begin{centering}
\includegraphics[scale=0.52]{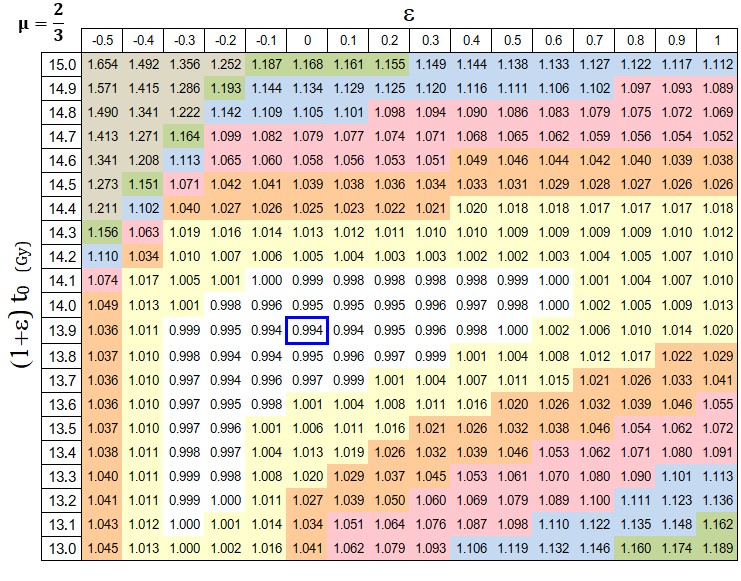}
\par\end{centering}
\begin{centering}
\includegraphics[scale=0.52]{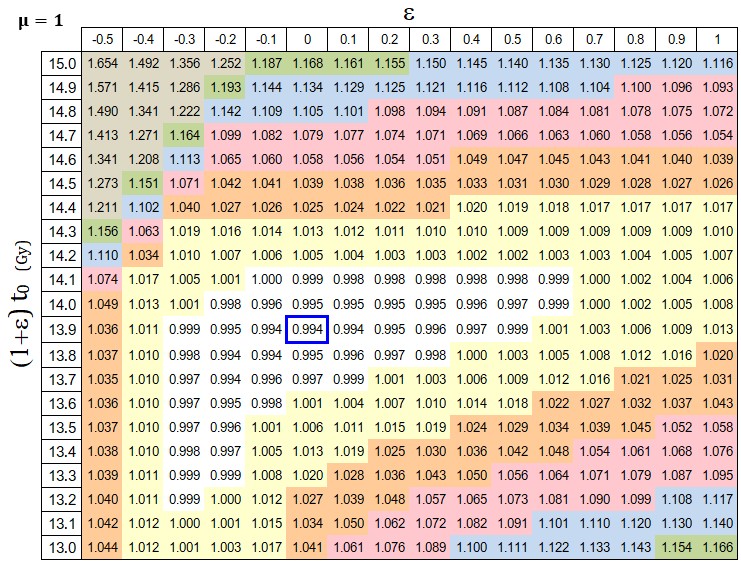}
\par\end{centering}
\noindent \caption{Value of $\chi$ on a 3-d grid. From top to bottom: $\mu=1/2,\ 2/3,\ 1$
resp. Stratifying color: $\chi<1$ in white; $\chi\in[1,1.02)$ in
yellow; $\chi\in[1.02,1.05)$ in orange; $\chi\in[1.05,1.1)$ in pink;
$\chi\in[1.1,1.15)$ in blue; $\chi\in[1.15,1.2)$ in green; $\chi\geqslant1.2$
in gray. The grid point with lowest $\chi$ is boxed in blue.}
\label{fig:3d-grid}
\end{figure}
\begin{figure*}
\includegraphics[scale=0.65]{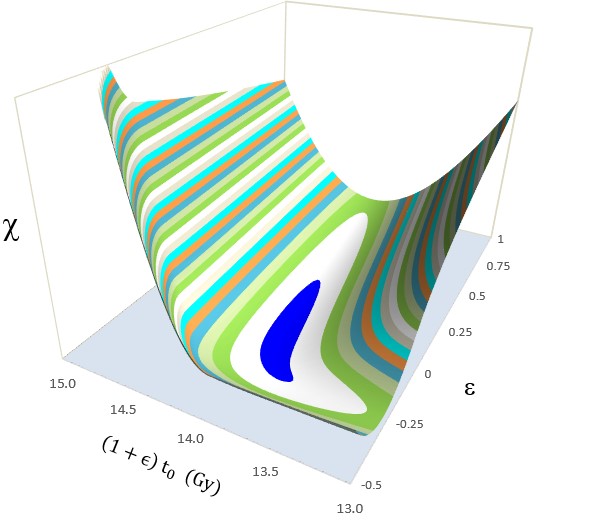}\includegraphics[scale=0.55]{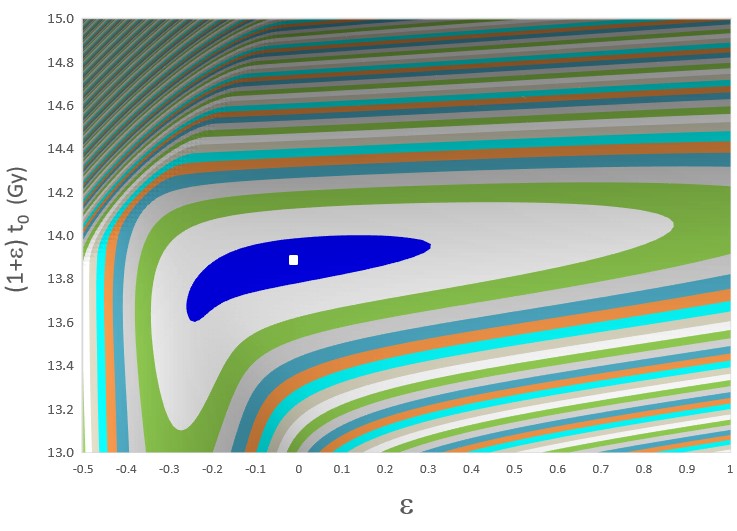}

\caption{Contour plots of $\chi$ for Formula \eqref{eq:simplified-formula}
in fine resolution. The dark blue area has lowest $\chi$ (smaller
than $0.995$). Each successive surrounding band has increasing $\chi$
with width $\Delta\chi=0.005$. The white dot in the right panel represents
$\chi_{min}\approx0.993942$, achieved at $t_{0}\approx14.2\text{ Gy},\ \epsilon=-0.02,\ F_{\infty}=0.9,\ \kappa=5.4$.}
\label{fig:simplified}
\end{figure*}

$ $

In \citep{Scolnic}, Scolnic and collaborators produced a dataset
of luminosity for $1,048$ objects with redshift $z$ ranging from
$0.01$ to $2.25$. Their actual Combined Pantheon Sample is accessible
in \citep{Pantheon-data}. The observational data are given in the
form of distance modulus $\mu:=m-M$ which is related to luminosity
distance $d_{L}$ as:
\begin{equation}
\mu:=m-M=5\log_{10}(d_{L}/Mpc)+25\label{eq:modulus}
\end{equation}
The Pantheon dataset provides, for each object $i^{th}$, the value
of redshift $z_{i}$, the value of $m_{i}$ together with its error
bar $\sigma_{i}$. For each data point $i^{th}$, we add in a constant
value of $\left|M\right|=19.35$ to recover the distance modulus $\mu_{i}^{\text{data}}$.

$ $

In Section \ref{subsec:Modifying-dL-z} we derived the modified luminosity
distance-redshift formula in Barrow's VSL context, Eq. \eqref{eq:modified-dL-vs-z},
and proposed a minimal model for the variation in the local scale
of gravitationally-bound regions, Eq. \eqref{eq:F-func}. The integral
in \eqref{eq:modified-dL-vs-z} must be carried out via numerical
means (tool available upon request). For each combination of $\{t_{0},\mu,\epsilon,F_{\infty},\kappa\}$
we use Formula \eqref{eq:modified-dL-vs-z} in conjunction with \eqref{eq:F-func}
to compute the VSL ``model'' luminosity distance $d_{L,i}^{[loc]\,\text{model}}$
for object $i^{th}$ then convert it to $\mu_{i}^{\text{model}}$
via Eq. \eqref{eq:modulus}. Our aim is to adjust the VSL model parameters
in order to minimize the $\chi^{2}$ of the difference between a model's
prediction $\left\{ \mu_{i}^{\text{model}}\right\} $ and the $n=1,048$
Pantheon data points $\left\{ \mu_{i}^{\text{data}}\right\} $ normalized
by the measurement error $\sigma_{i}$:
\begin{equation}
\chi^{2}:=\frac{1}{n}{\displaystyle \sum_{i=1}^{n}\frac{1}{\sigma_{i}^{2}}\left(\mu_{i}^{\mbox{data}}-\mu_{i}^{\mbox{model}}\right)^{2}}\label{eq:chi}
\end{equation}
Figure \ref{fig:3d-grid} is the result of our fit presented in a
3-dimensional grid $\{\mu,\epsilon,(1+\epsilon)t_{0}\}$. Note that
we arrange $t_{0}$ and $\epsilon$ in a combination $(1+\epsilon)t_{0}$.
The parameter $\mu$ takes in 3 values: $1/2,\,2/3,\,1$. The range
for $\epsilon$ is from $-0.5$ to $1$. The range for $(1+\epsilon)\,t_{0}$
is from $13$ Gy to $15$ Gy. For each point on the grid $\{\mu,\epsilon,(1+\epsilon)t_{0}\}$,
we minimize the $\chi^{2}$ error by adjusting the two parameters
$(F_{\infty},\kappa)$ of the function $F(z)$. Figure \ref{fig:3d-grid}
shows the value of $\chi$, with stratifying colors as explained in
the caption.

The fit shows a very weak dependence of $\chi$ on $\mu$. Across
the 3 tables in Fig. \ref{fig:3d-grid}, the variation in $\chi$
is negligible. Note that the column with $\epsilon=0$ is identical
for all 3 tables since for $\epsilon=0$ the parameter $\mu$ would
drop out of Formula \eqref{eq:modified-dL-vs-z}.

The central area with $\chi<1$ shown in white gathers around $(1+\epsilon)t_{0}\approx13.1-14.1$
Gy with $\epsilon$ ranging from $-0.3$ to $0.6$. The grid point
with lowest $\chi_{min}=0.99395$ indicated in the blue box occurs
at $\{\epsilon^{*},\,t_{0}^{*}\}=\{0,\,13.9\,\text{Gy}\}$ regardless
of $\mu$.

$ $

Given that the effect of $\mu$ is very weak, we also try with a simplified
version of \eqref{eq:modified-dL-vs-z} by suppressing $F(z)$ in
the integrand to have a reduced formula:
\begin{align}
\frac{d_{L}^{[loc]}}{c_{SS}} & =\frac{t_{0}(1+\epsilon)}{F(z)}\,(1+z)\,\int_{0}^{z}\frac{dz'}{(1+z')^{1+\epsilon}}\nonumber \\
 & =\frac{t_{0}(1+\epsilon)}{F(z)}\,(1+z)\,\frac{1}{\epsilon}\left[1-\frac{1}{(1+z)^{\epsilon}}\right]\label{eq:simplified-formula}
\end{align}
We repeat the fit of the Pantheon data using Formula \eqref{eq:simplified-formula}.
Figure \ref{fig:simplified} shows the contour plots for $\chi$ of
this exercise. The overall minimum of $\chi$ (shown by the white
dot in the right panel) is achieved at $t_{0}\approx14.2\text{ Gy},\ \epsilon=-0.02,\ F_{\infty}=0.9,\ \kappa=5.4$
which is very close to the vertical axis $\epsilon=0$.

\subsection{\label{subsec:Special-case-fit}Special case: Fitting for $\epsilon=0$}

\noindent The fit in Section \ref{subsec:General-fit} cannot reject
the null hypothesis of $\epsilon=0$. In what follows, we shall provide
a theoretical justification in support of $\epsilon=0$.

$ $

\noindent With the evolution rule \eqref{eq:evolution} of the cosmic
scale factor:
\begin{equation}
a\propto t^{\mu}
\end{equation}
on the dimensionality ground, the speed of light should vary as:
\begin{equation}
c\propto\frac{a}{t}\propto a^{1-\frac{1}{\mu}}
\end{equation}
Compared with \eqref{eq:Barrow-func}, $c\propto a^{-\zeta},$ we
have:
\begin{equation}
-\zeta=1-\frac{1}{\mu}
\end{equation}

\noindent or
\begin{equation}
\mu\,(1+\zeta)=1\label{eq:zeta-vs-mu}
\end{equation}
Combining \eqref{eq:zeta-vs-mu} with the definition of $\epsilon$
in \eqref{eq:definition-of-epsilon}, we arrive at:
\begin{equation}
\epsilon=0
\end{equation}

$ $

\begin{figure}
\noindent \begin{centering}
\includegraphics[scale=0.65]{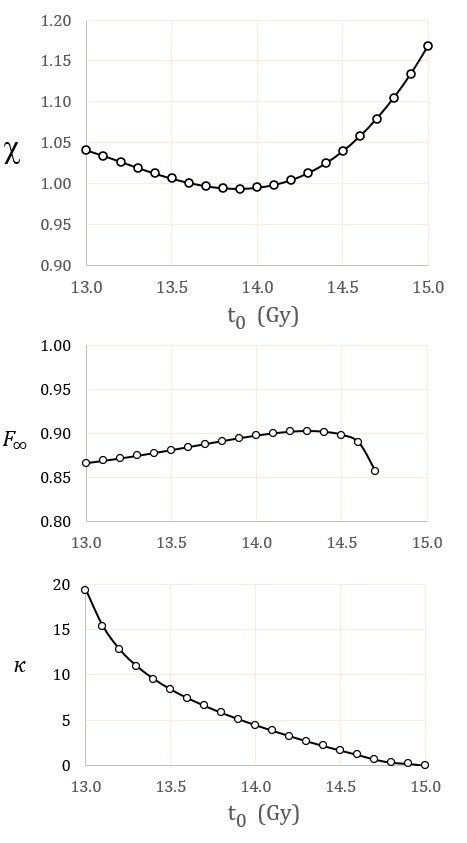}
\par\end{centering}
\caption{Special case: $\epsilon=0$. Optimal values of $\chi$, $F_{\infty}$,
and $\kappa$ as function of $t_{0}$. $\chi$ reaches minimum at
$t_{0}^{*}\approx13.9$ Gy.}

\label{fig:chi-epsilon-equal-0}
\end{figure}

\noindent In the rest of this subsection, we shall adopt $\epsilon=0$.
In this case, the luminosity distance-vs-\emph{z} \eqref{eq:modified-dL-vs-z}
is vastly simplified to:
\begin{equation}
\frac{d_{L}^{[loc]}}{c_{SS}}=\frac{t_{0}}{F(z)}\,(1+z)\,\ln(1+z)\label{eq:special-dL-vs-z}
\end{equation}
with $\mu$ dropping out of the equation. Figure \ref{fig:chi-epsilon-equal-0}
plots $\chi$ as a function of $t_{0}$, with $\chi$ achieving the
global minimum value of $\chi_{VSL}=0.99395$ at $t_{0}^{*}=13.9$
Gy, $F_{\infty}^{*}=0.894$, and $\kappa^{*}=0.502$. Also note that
in Figure \ref{fig:chi-epsilon-equal-0} for $t_{0}\gtrsim14.8$,
with $\kappa\rightarrow0$, by virtue of \eqref{eq:F-func}, $F(z)$
would approach $1$ for all $z$, rendering $F_{\infty}$ irrelevant.
We thus do not show $F_{\infty}$ in the right end of the middle plot.

$ $

\begin{figure}
\begin{centering}
\includegraphics[scale=0.8]{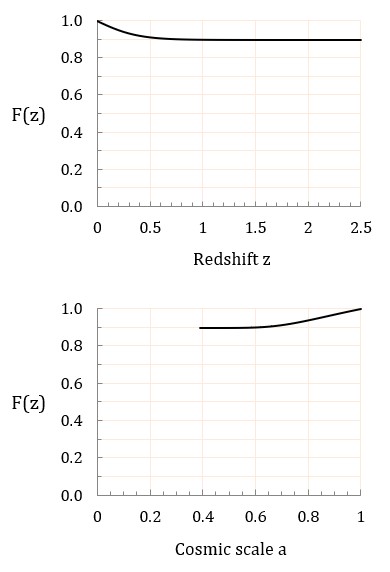}
\par\end{centering}
\caption{The variation of the local scale as functions of redshift (upper panel)
and of cosmic scale factor (lower panel).}

\label{fig:func_F}
\end{figure}

\begin{figure}
\begin{centering}
\includegraphics[scale=0.8]{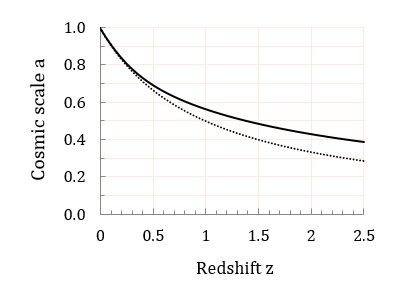}
\par\end{centering}
\caption{The variation of the global cosmic scale $a$ as a function of redshift.
Solid line: our VSL model with $\zeta=1/2$. Dotted line: classic
Lema\^itre's formula, viz. $1+z=a^{-1}$.}

\label{fig:a-vs-z}
\end{figure}

At the optimal point $\{t_{0}^{*},F_{\infty}^{*},\kappa^{*}\}$, the
variation in $F$ as a function of redshift $z$ in shown in the upper
panel of Figure \ref{fig:func_F}. Combining this knowledge of $F(z)$
with Eq. \eqref{eq:modified-Lemaitre-2}, we produce the variation
of $F$ as a function of the cosmic scale factor presented in the
lower panel of Figure \ref{fig:func_F} and the dependence of the
cosmic scale factor in terms of redshift displayed in Figure \ref{fig:a-vs-z}.
Also shown in Figure \ref{fig:a-vs-z} is the classic Lema\^itre
redshift formula.

\noindent 

\noindent 

\begin{figure}
\begin{centering}
\includegraphics[scale=0.8]{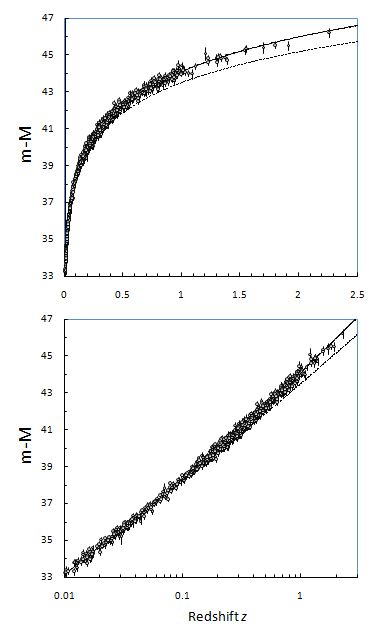}
\par\end{centering}
\caption{Comparison of various luminosity distance-redshift formulae fitted
to the Pantheon data. Open circles: 1,048 Pantheon data points with
error bars, listed in Ref. \citep{Pantheon-data}. Long-dashed line:
$\Lambda CDM$ model's formula \eqref{eq:LCDM-formula}, with $H_{0}=70.2,\,\Omega_{M}=0.285,\,\Omega_{\Lambda}=0.715,\,\Omega_{curv}=0$.
Dotted line: flat CDM model's formula \eqref{eq:flatCDM-formula},
with $H_{0}=70.2,\,\Omega_{M}=1,\,\Omega_{\Lambda}=0,\Omega_{curv}=0$.
Solid line: our VSL formulae \eqref{eq:special-dL-vs-z} and \eqref{eq:F-func}
with $t_{0}^{*}=13.9\text{ Gy},\ F_{\infty}^{*}=0.894,\ \kappa^{*}=5.202$.
Redshift shown in linear scale (upper panel) and in log scale (lower
panel).}

\label{fig:fit-w-F}
\end{figure}

\begin{figure}
\begin{centering}
\includegraphics[scale=0.8]{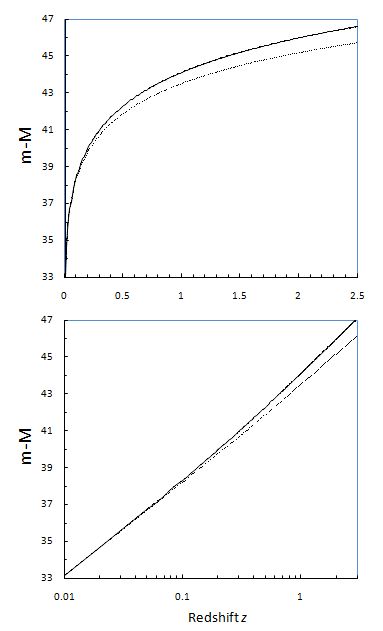}
\par\end{centering}
\caption{Same as Figure \ref{fig:fit-w-F} but with the Pantheon data points
removed for clarity. Our VSL model and the $\Lambda CDM$ model are
indistinguishable for all values of $z\in(0,2.25)$.}

\label{fig:fit-w-F-2}
\end{figure}

Figure \ref{fig:fit-w-F} displays the fits to our VSL model (solid
curve) and to the $\Lambda CDM$ (long-dashed curve), respectively.
The best fits produce $\chi_{VSL}=0.99395$ for our VSL model (Formula
\eqref{eq:special-dL-vs-z} with $t_{0}^{*}=13.9\text{ Gy},\ F_{\infty}^{*}=0.894,\ \kappa^{*}=5.202$)
and $\chi_{\Lambda CDM}=0.99410$ for $\Lambda CDM$ (Formula \eqref{eq:LCDM-formula}
with $H_{0}=70.2,\,\Omega_{M}=0.285,\,\Omega_{\Lambda}=0.715,\,\Omega_{curv}=0$).
The dotted curve corresponds the flat CDM model's Formula \eqref{eq:flatCDM-formula}
with $H_{0}=70.2$. Figure \ref{fig:fit-w-F-2} is identical to Figure
\ref{fig:fit-w-F} except that the Pantheon data points are removed
for clarity. It is worth commenting that our VSL fit (solid curve)
and the $\Lambda CDM$ fit (long-dashed curve) are \emph{indistinguishable}
in Figures \ref{fig:fit-w-F} and \ref{fig:fit-w-F-2}.

$ $

\begin{figure}
\noindent \begin{centering}
\includegraphics[scale=0.8]{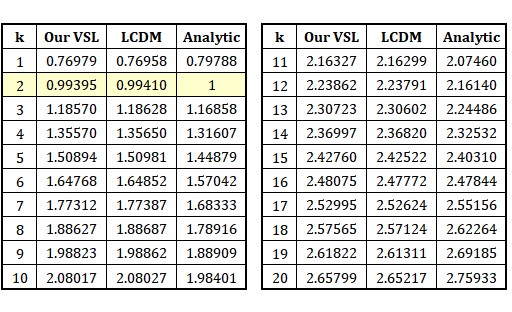}
\par\end{centering}
\caption{Goodness of fit for the absolute moments $L_{k}$ up to order $20^{th}$.
Definition of $L_{k}$ is in Eq. \eqref{eq:L_k-def}, with analytic
values given in Eq. \eqref{eq:L_k-analytic}.}

\label{fig:goodness_fit}
\end{figure}

To further probe the quality of the fits, we also compute the absolute
moments of the normalized errors, defined as:
\begin{equation}
L_{k}:=\left[\frac{1}{n}{\displaystyle \sum_{i=1}^{n}\frac{1}{\sigma_{i}^{k}}}\left|\mu_{i}^{\mbox{data}}-\mu_{i}^{\mbox{model}}\right|^{k}\right]^{1/k}\label{eq:L_k-def}
\end{equation}
which includes $\chi$ as a special case, $\chi\equiv L_{2}$. If
the normalized error $(\mu_{i}^{\text{data}}-\mu_{i}^{\text{model}})/\sigma_{i}$
follows a Gaussian distribution of zero mean and unit variance, the
analytical formula for its absolute moment of order $k$ is:
\begin{align}
L_{k} & =\left[\frac{1}{\sqrt{2\pi}}\int_{-\infty}^{\infty}dx\left|x\right|^{k}e^{-\frac{1}{2}x^{2}}\right]^{1/k}\nonumber \\
 & =\begin{cases}
\left[(k-1)!!\sqrt{\frac{2}{\pi}}\right]^{1/k}\text{ if k odd}\\
\ \ \ \ \left[(k-1)!!\right]^{1/k}\ \ \ \ \ \ \text{if k even}
\end{cases}\label{eq:L_k-analytic}
\end{align}
Figure \ref{fig:goodness_fit} lists the absolute moments of our VSL,
the $\Lambda CDM$, and the Gaussian analytic. Our VSL model is indistinguishable
from the $\Lambda CDM$ in terms of the absolute moments up to the
order $20^{th}$.

\noindent 

$ $

The meaning and implications of $F(z)$ will be discussed in Section
\ref{subsec:Role-of-local-scale}. Although our VSL fit involves 3
adjustable parameters, the meaning of the 2 parameters ($F_{\infty}$
and $\kappa$) is intuitive and direct: they are to account for a
variation of the local scale in the gravitationally-bound regions
over the course of cosmic time.

\section{\label{sec:Interpretations-of-results}Interpretations of results}

\noindent There are 2 agents at play in our VSL approach:\renewcommand{\labelenumi}{\#\arabic{enumi})}
\begin{enumerate}
\item A variation in the velocity of light, characterized by the parameter
$\zeta$ (or equivalently $\mu$ in the relation $1+\zeta=1/\mu$);
\item A variation in the local scale of gravitationally-bound regions, parameterized
by $F_{\infty}$ and $\kappa$.
\end{enumerate}
\renewcommand{\labelenumi}{\arabic{enumi}.}We discuss the meaning
and implications of each agent in this Section.

\subsection{\label{subsec:New-interpretation}A new interpretation of the acceleration
in Type Ia SNe by way of VSL}

\noindent 
\begin{figure}
\begin{centering}
\includegraphics[scale=0.8]{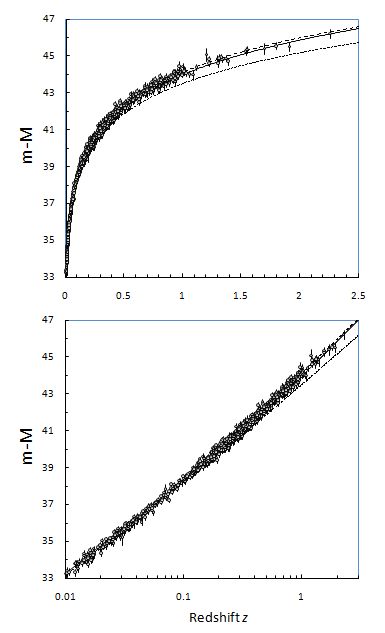}
\par\end{centering}
\caption{Fitting the Pantheon dataset with Formula \eqref{eq:without-F} which
deactivates the function $F(z)$. Open circles: 1,048 data points
with error bars, listed in Ref. \citep{Pantheon-data}. Solid line:
VSL Formula \eqref{eq:without-F} with $t_{0}=14.7$ Gy. Long-dashed
line: $\Lambda$CDM model Formula \eqref{eq:LCDM-formula} with $H_{0}=70.2,\,\Omega_{M}=0.284,\,\Omega_{\Lambda}=0.716$.
Dotted line: flat CDM model Formula \eqref{eq:flatCDM-formula} with
$H_{0}=70.2$. Redshift shown in linear scale (upper panel) and in
log scale (lower panel).}

\label{fig:fit-without-F}
\end{figure}

\noindent We first focus on Agent \#1. To do so, we disable Agent
\#2 by suppressing the variation in the local scale $F(z)$ in Eq.
\eqref{eq:special-dL-vs-z} to obtain:
\begin{equation}
\frac{d_{L}^{[loc]}}{c_{SS}}=t_{0}\,(1+z)\,\ln(1+z)\label{eq:without-F}
\end{equation}
We then carry out a refit of the Pantheon dataset to Formula \eqref{eq:without-F}
which has only 1 free parameter, $t_{0}$. Note that $\zeta$ is absent
from this Formula. The result of the fit is displayed in Figure \ref{fig:fit-without-F}.
The 1,048 Pantheon data points are shown in open circles with error
bars. The solid line shows Formula \eqref{eq:without-F} corresponding
to one parameter $t_{0}=14.7$ Gy with $\chi_{min}=1.1197$. The long-dashed
line corresponds to $\Lambda CDM$ model's Formula \eqref{eq:LCDM-formula}
with 3 parameters $H_{0}=70.2,\,\Omega_{M}=0.284,\,\Omega_{\Lambda}=0.716,\,\Omega_{curv}=0$
with $\chi_{min}=0.9941$. The dotted line shows the flat CDM model's
Formula \eqref{eq:flatCDM-formula} with $H_{0}=70.2$.

$ $

The most outstanding feature in Figure \ref{fig:fit-without-F} is
that in the high-$z$ section the solid curve (i.e., VSL model) with
only 1 single adjustable parameter $t_{0}$ acquires an upward slopping
above the dotted curve (i.e., the flat CDM model). This upward slopping
behavior of SNeIa has been interpreted as the hallmark of an acceleration
in the recent epoch. Overall, the solid curve tracks the SNe data
reasonably well; it closely resembles the long-dashed curve (i.e.,
the $\Lambda CDM$ model) which has 3 adjustable parameters $(H_{0},\,\Omega_{M},\,\Omega_{\Lambda})$.
Also note that this ``optimal'' value of $t_{0}=14.7$ Gy is not
too far off from the value of $t_{0}^{*}=13.9$ Gy obtained via Formula
\eqref{eq:special-dL-vs-z} which activates the use of $F(z)$. Thus,
even without allowing the variation in the local scale of gravitationally-bound
regions, the variation in the velocity of light \emph{alone} is already
able to capture the behavior of the SNe data in the Pantheon Sample.
In particular, the distance modulus $\mu$ for the high-$z$ supernovae
exceeds what would have been expected from the flat CDM model with
$H_{0}=70.2$ and tracks the $\Lambda CDM$ model quite closely in
both low-\emph{z} and high\emph{-z} sections. Whereas standard cosmology
and the VSL approach each is able to account for the excess in $\mu$
compared with the flat CDM model in the high-z section, standard cosmology
needs to resort to the cosmological component $\Omega_{\Lambda}$,
whereas the VSL approach bypasses $\Omega_{\Lambda}$.

$ $

Parsimony asides, let us seek the intuition behind the accelerating
expansion in Type Ia SNe in the light of our VSL-based analysis. In
our analysis, the Pantheon dataset is found to be consistent with
a family of universes. In each member in the family, the mode of expansion
is $a\propto t^{\mu}$ and the light speed varies in the fashion $c\propto a^{-\zeta}$
with $\zeta=1-1/\mu$ as dictated by Eq. \eqref{eq:zeta-vs-mu}. By
itself, the Pantheon dataset is silent about the value of $\mu$ (and
$\zeta)$. We shall consider 2 most relevant modes of expansion in
what follows.

\subsubsection*{The linear expansion mode, $a\propto t$}

\noindent This universe, with $\mu=1$, corresponds to $\zeta=0$;
namely, the speed of light is unchanged over the course of its expansion.
In other words, the Pantheon data is consistent with a universe with
linear expansion and a non-variable speed of light. In the later epoch,
this mode of expansion is faster than the critical model in the flat
CDM model, $a\propto t^{2/3}$. This means that, compared with the
baseline case of $a\propto t^{2/3}$, this universe does accelerate.

\subsubsection*{The critical expansion mode, $a\propto t^{2/3}$}

\noindent This universe, with $\mu=2/3$, corresponds to $\zeta=1/2$;
namely, the speed of light decreases as the universe expands, $c\propto a^{-1/2}$.
Let us deep dive into this very interesting case. 

$ $

Consider two supernovae A and B at distances $d_{A}$ and $d_{B}$
away from the Earth such that $d_{B}=2\,d_{A}$. In standard cosmology,
their redshift values $z_{A}$ and $z_{B}$ are related: $z_{A}\approx2\,z_{A}$
(to the first-order approximation). However, this relation breaks
down in the VSL context. In the VSL universe which accommodates the
variation in the speed of light in the form�$c\propto a^{-1/2}$,
light had traveled faster in the distant past (when the cosmic factor
$a\ll1$) than it did in the more recent epoch (when $a\lesssim1$).
Therefore, the photon emitted from supernova B could cover twice as
long the distance in less than twice the amount of time as would be
required from the photon emitted from supernova A. Having spent less
time in transit than what standard cosmology would have demanded,
the B-photon experienced less cosmic expansion than expected, and
thus experienced a lower redshift than the classic Lema\^itre formula
would have required. Namely:
\begin{equation}
z_{B}<2\,z_{A}\ \ \text{for}\ \ d_{B}=2\,d_{A}
\end{equation}
Conversely, consider a supernova C with $z_{C}=2\,z_{A}$. In order
for the C-photon to have experienced twice as much the redshift as
the A-photon did, the C-photon must travel a distance exceeding twice
as long compared with the A-photon: $d_{C}>2\,d_{A}$. This is because
since the C-photon traveled faster at the beginning of its journey
toward Earth, it must start at a farther distance (thus appearing
fainter than expected) to have experienced enough cosmic expansion
and hence the redshift. Namely:
\begin{equation}
d_{C}>2\,d_{A}\ \ \text{for}\ \ ~z_{C}=2\,z_{A}
\end{equation}
Therefore, the distance-vs-$z$ plot gains an additional upward
slope in the high-z section, as is captured in the behavior of the
solid curve in Figure \ref{fig:fit-without-F}. We conclude:

$ $

\noindent \emph{For a universe which expands in the critical mode,
$a\propto t^{2/3}$, the acceleration is equivalent to a variation
in the velocity of light in the $c\propto a^{-1/2}$ fashion.}

$ $

Recall that in 1911 during his search for a formulation of GR, Einstein
originated the possibility of VSL \citep{Einstein1911,Einstein1912-1,Einstein1912-2}.
He explicitly allowed the gravitational field to influence the \emph{value}
of the velocity of light\emph{; }see Page 903 of Ref. \citep{Einstein1911}:\emph{
``If $c_{0}$ denotes the velocity of light at the coordinate origin,
then the velocity of light $c$ at a point with a gravitational potential
$\Phi$ will be given by the relation: $c=c_{0}\left(1+\Phi/c^{2}\right)$''}.
In Refs. \citep{Einstein1912-1,Einstein1912-2}, Einstein further
emphasized the limited scope of the principle of the constancy of
$c$: the constancy of $c$ (and equivalently the Michelson-Morley
experimental result and the Lorentz invariance) are valid only \emph{locally},
and thus the variation in $c$ is not in contradiction with the constancy
of $c$. Although Einstein eventually did not incorporate his VSL
idea into his final 1915-16 GR theory in favor of the Riemannian manifold,
his insight of a location-dependent velocity of light is a legitimate
pursuit on its own right and merit. While the geometrical approach
proved successful in accounting for the Solar System Phenomenology
at Einstein's time, the possibility of VSL could yet provide crucial
hints into the challenges that scientists currently encounter beyond
the solar system; in particular, the nature of the $\Omega_{\Lambda}$
component in the $\Lambda CDM$ model and that of the acceleration
discovered in Type Ia supernovae.

$ $

The Friedmann equations are based on GR and extrapolate the theory
into the cosmic domain. This extrapolation embodies a major assumption
in cosmology, as summarized in the Review {[}20{]}. Despite GR's successes
in accounting for the Solar System Phenomenology,\emph{ is it safe
to extrapolate GR into the cosmic domain without prudently considering
Einstein's 1911 idea regarding VSL \citep{Einstein1911,Einstein1912-1,Einstein1912-2}?}
While negligible in the solar system, on the cosmic scale, the impact
of variation in $c$ can accumulate, and as our analysis
expounded, may manifest in the accelerating expansion that is observed
in Type Ia supernovae.

$ $

If the critical expansion mode can be justified theoretically, the
ground-breaking `1998 discoveries of the acceleration \citep{Riess1,Perlmutter}
could qualify as evidence in support of a decreasing velocity of light
as the universe expands. \footnote{In our follow-up report \citep{Nguyen-theoretical}, we shall provide
the theoretical basis in support of the critical expansion mode, $a\propto t^{2/3}$,
and the variation rule $c\propto a^{-1/2}$ for the VSL universe.}

\subsection{\label{subsec:Role-of-local-scale}Role of local scale of gravitationally-bound
regions in the estimate of the Hubble constant}

\begin{figure}
\noindent \begin{centering}
\includegraphics[scale=0.7]{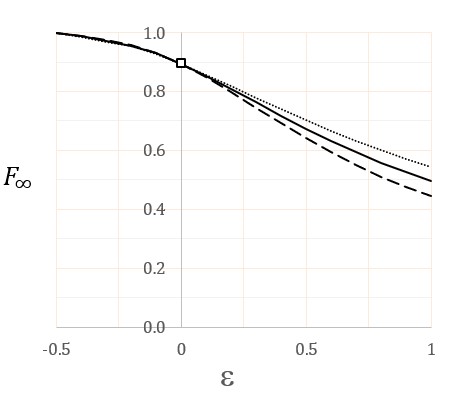}
\par\end{centering}
\caption{$F_{\infty}$ as function of $\epsilon$. Solid line: $\mu=2/3$;
dashed line: $\mu=1/2$; dotted line: $\mu=1$. All 3 lines cross
at $\{\epsilon^{*},F_{\infty}^{*}\}=\{0,\,0.894\}$ shown by the opened
square.}

\label{fig:F_infty}
\end{figure}

\noindent Let us turn to Agent \#2. The full consideration presented
in Section \ref{subsec:General-fit} was based on Formula \eqref{eq:modified-dL-vs-z}
in conjunction with \eqref{eq:F-func}. Given the values of the pair
$(\mu,\,\epsilon)$, we found the ``optimal'' combination of $\{t_{0},F_{\infty},\kappa\}$
which minimizes $\chi^{2}$, defined in \eqref{eq:chi}. The result
for the ``optimal'' $F_{\infty}$ is shown in Fig. \ref{fig:F_infty}
for $\mu=1/2,\,2/3,\,1$ and $\epsilon$ in the range $[-0.5,\,1]$.
At $\epsilon=0$, the luminosity distance-redshift formula \eqref{eq:modified-dL-vs-z}
is independent of $\mu$. Thus, in Fig. \ref{fig:F_infty}, all 3
curves thus cross at $F_{\infty}^{*}\approx0.9$ and $\epsilon=0$
(whereas $\kappa^{*}\approx5.2$ and $t_{0}^{*}\approx13.9$ Gy).

$ $

Although we only invoked a parsimonious function, Eq. \eqref{eq:F-func},
to model the variation in the local scale, the result obtained in
Figure \ref{fig:F_infty} is intuitive and promising. For a wide range
of $\epsilon$, the value $F_{\infty}$ is smaller than 1, meaning
that the high-\emph{z} emitters corresponded to a smaller local scale.
With $F_{\infty}^{*}\approx0.9$ and $\kappa^{*}\approx5.2$, Figure
\ref{fig:func_F} depicts a monotonic increase of the local scale
as the universe expands. It indicates that the gravitationally-bound
regions \emph{cannot} fully resist cosmic expansion. As the global
scale expanded, the galaxies also expanded by about 10\% larger over
the course of time since the recombination event, with the crossover
occurring at $z^{*}=1/\kappa^{*}\approx0.2$. High-z gravitationally-bound
regions that contained distant supernovae thus had a local scale which
is 10\% smaller than the local scale of our Solar System at our current
time.

$ $

An interesting question naturally arises: Considering that the Cosmic
Microwave Background (CMB) corresponded to a region of very high $z$
($\sim1,100$), should the CMB analysis take into account the 10-percent
reduction in the local scale for high-$z$ regions, as encoded in
the value of $F_{\infty}^{*}\approx0.9$?

$ $

This is a tantalizing possibility: the 10-percent reduction in the
local scale for early-time emitters uncovered in our analysis curiously
happens to be of similar magnitude as the discrepancy in the value
of $H_{0}$ when extracted from the CMB ($H_{0}\approx67$) \citep{tension4,tension5,tension6,tension7,tension8,tension9,tension10,tension11,tension12,tension13}
versus the one obtained from late-time objects ($H_{0}\approx74$)
\citep{tension1,tension2,tension3}, a 9\% difference. It would be
interesting to see whether the tension in the estimates of $H_{0}$
may be rooted in the variation of the local scale.

$ $

In what follows, we would suggest for such a potential connection.
With the aid of the age formula \eqref{eq:age-formula}, we convert
Eq. \eqref{eq:special-dL-vs-z} into:
\begin{equation}
\frac{d_{L}^{[loc]}}{c_{SS}}=\frac{\mu}{F(z)\,H_{0}}\,(1+z)\,\ln(1+z)\label{eq:special-dL-vs-z-2}
\end{equation}
The ``effective'' value of the Hubble constant for a given $z$
can be read off as $H_{0}^{\text{eff}}(z):=F(z)\,H_{0}$. With $F(z\rightarrow0)=1$
and $F(z\rightarrow\infty)=F_{\infty}^{*}$, the ``effective'' Hubble
constant is:
\begin{equation}
H_{0}^{\text{eff}}(z)=\begin{cases}
\begin{array}{c}
H_{0}\\
F_{\infty}^{*}H_{0}
\end{array} & \begin{array}{c}
\text{for }z\rightarrow0\\
\text{for }z\rightarrow\infty
\end{array}\end{cases}
\end{equation}
in which $F_{\infty}^{*}\approx0.9$ as obtained from our VSL fit
to the Pantheon data.

$ $

This probable ``explanation'' of the Hubble constant tension via
the use of $H_{0}^{\text{eff}}(z)$ can be further verified in a simple
way. Let us split the Pantheon dataset into 4 quartiles of equal size,
each containing 262 supernovae with redshift ranging from the lowest
to the highest. The $1^{st}$ quartile corresponds to $z\in(0.01,\,0.13)$;
the $4^{th}$ quartile to $z\in(0.4235,\,2.26)$. Next, let us fit
the $1^{1st}$ quartile and the $4^{th}$ quartile \emph{separately}
to the following formula:
\begin{equation}
\frac{d_{L}^{[loc]}}{c_{SS}}=t_{0}\,(1+z)\,\ln(1+z)\label{eq:tmp-9}
\end{equation}
This formula is nothing but \eqref{eq:special-dL-vs-z-2} with $t_{0}$
standing in place of the prefactor $\mu/(F(z)\,H_{0})$; $\ t_{0}$
thus acts as an ``effective'' parameter that is valid for a given
quartile of the dataset. The results of the fit are:
\begin{itemize}
\item The $1^{st}$ quartile (lowest $z$) yields $t_{0}^{\text{[Q1]}}=14.1$
Gy;
\item The $4^{th}$ quartile (highest $z$) yields $t_{0}^{\text{[Q4]}}=15.4$
Gy.
\end{itemize}
There is indeed a $9\%$ difference between $t_{0}^{\text{[Q1]}}$
and $t_{0}^{\text{[Q4]}}$, thus lending support to our tentative
``explanation'' of the Hubble constant tension.

\newpage{}

\section{\label{sec:summary}Summary}

\noindent We applied a version of VSL to late-time cosmology. In our
consideration, the velocity of light varies as a function of the scale
factor in the form $c\propto a^{-\zeta}$ , first put forth by Barrow
in \citep{Barrow3}. There is one key departure in our approach however.
In \citep{Barrow3} the scale factor was intended to mean the global
cosmic scale factor. In our approach, the scale factor is also applicable
to the local scale factor of the gravitationally-bound regions (namely,
galaxies) as well. Densely populated regions (such as galaxies) which
contain the emitter sources (e.g., a supernova) or the observer (in
this case, the dense region is the Milky Way) are gravitationally-bound
and are not subject to cosmic expansion. On the one hand, this aspect
is crucial in order for the Earth-based observer being able to detect
the redshift. On the other hand, this aspect is a great virtue: Since
the Milky Way has resisted the cosmic expansion, it has a different
scale from the global cosmic scale of the outer space region that
directly encloses it. Thanks to the VSL dependence $c\propto a^{-\zeta}$,
the velocity of light inside the Milky Way must \emph{differ} from
that in the outer space region that directly encloses it. For a lightwave
from the outer space entering the Milky Way to reach the Earth-based
astronomer, the change in the speed of light at this juncture would
force the lightwave to undergo a refraction. This refraction effect
alters the wavelength that would eventually reach the astronomer's
apparatus.

$ $

The refraction effect is what was missing in previous works that attempted
to apply VSL to the observational data \citep{Zhang,Qi,Ravanpak,Salzano}.
Without including the refraction effect, one would continue to use
the classic Lema\^itre redshift formula $1+z=a^{-1}$. However, by
properly taking into account the refraction effect, we find that the
standard Lema\^itre redshift formula needs be replaced by the \emph{modified}
Lema\^itre redshift formula $1+z=a^{-(1+\zeta)}$. The participation
of the VSL exponent $\zeta$ in the modified Lema\^itre redshift
formula is novel; it holds the key to our re-analysis of the Combined
Pantheon Sample in the context of VSL. To the best of our knowledge,
our current work is the first to bring the refraction effect to the
fore.

$ $

In our reformulation of the distance-redshift relations, apart from
the RW metric (with $c$ being allowed to vary), we do \emph{not}
rely on any underlying theory of gravitation, such as GR or the Friedmann
equations. Our analysis is applicable to a universe which satisfies
2 following parsimonious conditions:

\noindent \renewcommand{\labelenumi}{(\Roman{enumi})}
\begin{enumerate}
\item A variation in the velocity of light as a power-law function of the
scale factor: $c\propto a^{-\zeta}$. This functional form does not
have a preferred scale.
\item An evolution for the global cosmic scale factor as a power-law function
of ``cosmic time'': $a\propto t^{\mu}$. This functional form does
not have a preferred time scale. The flat CDM model is a member (when
$\mu=2/3$) of this family.
\end{enumerate}
\renewcommand{\labelenumi}{\arabic{enumi}.}

$ $

\noindent The flow of exposition in our work was as follows:
\begin{enumerate}
\item Modifying Lema\^itre's redshift formula: We demonstrated how the
VSL rule (I), $c\propto a^{-\zeta}$, gives rise to the refraction
effect. We highlighted the role of the 3 transits between different
regions as a lightwave traveling from a distant emitter to reach the
Earth-based astronomer. See Sections \ref{subsec:Role-grav-bound}
and \ref{subsec:Refraction-effect}. The cumulative effects of the
3 transits lead to a new formula for the redshift:
\begin{equation}
1+z=a^{-(1+\zeta)}\,F^{1+\zeta}(z)\label{eq:summary-1}
\end{equation}
in which the function $F(z)$ captures the variation in the local
scale of gravitationally-bound regions as a function of the redshift.
See Section \ref{subsec:Modifying-Lemaitre-formula}.
\item Modifying the distance-vs-$z$ formula: Employing the evolution rule
(II), $a\simeq t^{\mu}$, we derived a new distance-vs-\emph{z} relation:
\begin{equation}
\frac{r}{c_{0}}=\frac{1}{(1+\zeta)\,H_{0}}\int_{0}^{z}dz'\,\frac{F^{\epsilon\,(1+\zeta)}(z')}{(1+z')^{1+\epsilon}}\label{eq:summary-2}
\end{equation}
in which $1+\epsilon:=\frac{1}{\mu(1+\zeta)}$. See Section \ref{subsec:Modifying-d-z}.
\item Modifying the luminosity distance-vs-\emph{z} formula, which is the
centerpiece of our study:
\begin{equation}
\frac{d_{L}^{[loc]}}{c_{SS}}=\frac{t_{0}\,(1+\epsilon)}{F(z)}(1+z)\int_{0}^{z}dz'\,\frac{F^{\frac{\epsilon}{\mu(1+\epsilon)}}(z')}{(1+z')^{1+\epsilon}}\label{eq:summary-3}
\end{equation}
We further specified the variation of the local scale as a monotonic
slowly-varying function of $z$:
\begin{equation}
F(z):=F_{\infty}+(1-F_{\infty})\frac{2}{1+e^{\kappa z}}\label{eq:summary-4}
\end{equation}
See Section \ref{subsec:Modifying-dL-z}.
\item General analysis of the Pantheon dataset based on our reformulation
of redshift formulae: We applied Formulae \eqref{eq:summary-3} and
\eqref{eq:summary-4} to the Combined Pantheon Sample. We reported
the $\chi^{2}$ error between our VSL model and the Pantheon data
for a $3$-d grid $\{\mu,\epsilon,(1+\epsilon)t_{0}\}$. The $\chi^{2}$
error is found to be insensitive to the value of $\mu$. Moreover,
the minimum $\chi^{2}$ is consistent with $\epsilon=0$. See Section
\ref{subsec:General-fit}.
\item Special case, $\epsilon=0$: Based on a generic dimensionality argument,
we established a link between $\zeta$ and $\mu$: $\zeta=1-1/\mu$,
hence justifying $\epsilon=0$. The luminosity distance-vs-$z$ relation
is significantly simplified to:
\begin{equation}
\frac{d_{L}^{[loc]}}{c_{SS}}=\frac{t_{0}}{F(z)}\,(1+z)\,\ln(1+z)\label{eq:summary-5}
\end{equation}
It achieves an excellent fit to the Pantheon data with the cosmic
age parameter $t_{0}^{*}\approx13.9$ Gy; and $F_{\infty}^{*}\approx0.9$
and $\kappa^{*}\approx5.2$ for the function $F(z)$. The fit is \emph{indistinguishable}
from the $\Lambda CDM$ model; the absolute moment of the normalized
error term $L_{k}:=\left\langle \left|(\text{\ensuremath{\mu^{\text{Pantheon}}}-\ensuremath{\mu^{\text{model}}}})/\sigma\right|^{k}\right\rangle ^{1/k}$
is almost identical in both models (VSL and $\Lambda CDM$) for all
orders $k$ up to $20$. See Section \ref{subsec:Special-case-fit}.
\end{enumerate}
\noindent The parameters in Formulae \eqref{eq:summary-4} and \eqref{eq:summary-5}
are intuitive and have direct physical meanings. In particular:
\begin{itemize}
\item The cosmic age $t_{0}^{*}\approx13.9$ Gy is obtained, without invoking
the $\Lambda$ component.
\item The value of $F_{\infty}^{*}\approx0.9$ indicates a 10\% reduction
in the ``effective'' value of $H_{0}$ if the latter is estimated
from high-$z$ portion of the Pantheon dataset, as compared with the
value of $H_{0}$ estimated from low-$z$ portion of the Pantheon
dataset.
\end{itemize}
Two important implications emerge from our analysis:

\noindent \renewcommand{\labelenumi}{[\Alph{enumi}]}
\begin{enumerate}
\item We offer a viable interpretation of the accelerating expansion on
the sole basis of variable speed of light, in place of the $\Lambda$
component. See Section \ref{subsec:New-interpretation}.
\item As a by-product, the role of $F(z)$ might shed a refreshing perspective
onto the ongoing tension in the estimate of the Hubble constant using
the CMB versus that derived from late-time objects \citep{tension1,tension2,tension3,tension4,tension5,tension6,tension7,tension9,tension8,tension10,tension11,tension12,tension13}.
See Section \ref{subsec:Role-of-local-scale}.
\end{enumerate}
\renewcommand{\labelenumi}{\arabic{enumi}.}

\noindent In conclusion, the Pantheon SNeIa dataset is consistent
with a variable-light-speed scenario in which $a\propto t^{\mu}$
and $c\propto a^{1-1/\mu}$ with $\mu$ left \emph{unspecified}. In
a universe which expands as $a\propto t^{2/3}$, the velocity of light
would vary as $c\propto a^{-1/2}$.

$ $

\noindent A theoretical basis for the VSL form $a\propto t^{2/3}$
and $c\propto a^{-1/2}$ will be provided in our follow-up report,
Ref. \citep{Nguyen-theoretical}.
\begin{acknowledgments}
\noindent The idea of applying VSL in Barrow's form, $c\propto a^{-\zeta}$,
to the Pantheon dataset sprang out of the author's communications
with Viktor Toth, who also generously offered us insightful theoretical
guidance during the development of our work. The author thanks Richard
Shurtleff and Vesselin Gueorguiev for their constructive feedbacks;
and Richard Shurtleff, Juha Korpela, and Lucas Lombriser for their
meaningful encouragements.

Our numerical tool (1Mb) and the output of our analysis presented
in Section \ref{sec:Fitting-of-modified-formulae} are available upon
request.
\end{acknowledgments}

\appendix

\section{\label{sec:Equivalent-derivation}An equivalent derivation of the
modified Lema\^itre redshift formula}

\noindent We produce an alternative route by way of frequency transformation
to modifying Lema\^itre's redshift formula \eqref{eq:modified-Lemaitre-1}.
The modified RW metric \eqref{eq:modified-RW-zeta} can be recast
as:
\begin{equation}
ds^{2}=a^{2}\left(t\right)\,\left[\frac{c_{0}^{2}}{a^{2+2\zeta}\left(t\right)}dt^{2}-\frac{dr^{2}}{1-kr^{2}}-r^{2}d\Omega^{2}\right]
\end{equation}
The null geodesic ($ds^{2}=0$) for a lightwave traveling from an
emitter toward Earth (viz. $d\Omega=0$) is thus:
\begin{equation}
\frac{c_{0}\,dt}{a^{1+\zeta}(t)}=\frac{dr}{\sqrt{1-kr^{2}}}\label{eq:null_geodesic}
\end{equation}
Denote $t_{e}$ and $t_{o}$ the emission and observation time points
of the lightwave, and $r_{e}$ the co-moving distance of the galaxy
from Earth. From \eqref{eq:null_geodesic}, we have:
\begin{equation}
\int_{t_{e}}^{t_{o}}\frac{c_{0}\,dt}{a^{1+\zeta}(t)}=\int_{r_{e}}^{0}\frac{dr}{\sqrt{1-kr^{2}}}
\end{equation}
The next wavecrest to leave the emitter at $t_{e}+\delta t_{e}$ and
arrive at Earth at $t_{o}+\delta t_{o}$ satisfies:
\begin{equation}
\int_{t_{e}+\delta t_{e}}^{t_{o}+\delta t_{o}}\frac{c_{0}\,dt}{a^{1+\zeta}(t)}=\int_{r_{e}}^{0}\frac{dr}{\sqrt{1-kr^{2}}}
\end{equation}
Subtracting the two equations yields:
\begin{equation}
\frac{\delta t_{o}}{a^{1+\zeta}(t_{o})}=\frac{\delta t_{e}}{a^{1+\zeta}(t_{e})}
\end{equation}
which leads to the ratio between the emitted frequency and the observed
frequency: \footnote{\noindent In previous VSL analyses \citep{Zhang,Qi,Ravanpak}, it
was correctly reported, due to Eq. \eqref{eq:freq_shift}, that
\begin{equation}
\frac{\lambda_{o}}{\lambda_{e}}=\frac{c_{o}/\nu_{o}}{c_{e}/\nu_{e}}=\frac{c_{o}}{c_{e}}\frac{\nu_{e}}{\nu_{o}}=\frac{a_{o}^{-\zeta}}{a_{e}^{-\zeta}}.\frac{a_{o}^{1+\zeta}}{a_{e}^{1+\zeta}}=\frac{a_{o}}{a_{e}}
\end{equation}
which is identical to the classic consideration of non-variable speed
of light. However, that fact alone is not sufficient to conclude that
the classic Lema\^itre redshift formula remains valid for VSL. The
reason is that $\lambda_{o}$ is \emph{not} what the Earth-based astronomer
observes. To reach the astronomer, the lightwave needs to enter the
\emph{gravitationally-bound} Milky Way which has a scale \emph{smaller}
than the current global cosmic scale. Per $c\propto a^{-\zeta}$,
the velocity of light $c_{o}^{[loc]}$ inside the Milky Way is different
from the velocity of light $c_{o}$ in the outer space region that
encloses the Milky Way. The lightwave thus gets refracted during its
entry to the Milky Way, with its wavelength getting altered to $\lambda_{o}^{[loc]}=\lambda_{o}\,c_{o}^{[loc]}/c_{o}$
per Eq. \eqref{eq:Transit3}. It is the wavelength $\lambda_{o}^{[loc]}$
which gets registered in the astronomer's apparatus.}
\begin{equation}
\frac{\nu_{o}}{\nu_{e}}=\frac{\delta t_{e}}{\delta t_{o}}=\frac{a^{1+\zeta}(t_{e})}{a^{1+\zeta}(t_{o})}=\frac{a_{e}^{1+\zeta}}{a_{o}^{1+\zeta}}\label{eq:freq_shift}
\end{equation}

\noindent For transits between local regions to global regions (i.e.,
Transit \#1 and Transit \#3 in Fig. \ref{fig:wavetrain-full} in Page
\pageref{fig:wavetrain-full}), since $\lambda\propto c$, the frequency
is:
\begin{align}
\nu & =\frac{c}{\lambda}=\text{const }
\end{align}
This means that the frequency of the lightwave does not change during
Transit \#1 and Transit \#3, viz.
\begin{align}
\nu_{e}^{[loc]} & =\nu_{e}\\
\nu_{o}^{[loc]} & =\nu_{o}
\end{align}
Given that
\begin{align}
\lambda_{o}^{[loc]} & =\frac{c_{o}^{[loc]}}{\nu_{o}^{[loc]}}\\
\lambda_{e}^{[loc]} & =\frac{c_{e}^{[loc]}}{\nu_{e}^{[loc]}}\\
\frac{\lambda_{o}^{*}}{\lambda_{e}^{[loc]}} & =\frac{a_{o}^{[loc]}}{a_{e}^{[loc]}}
\end{align}
\begin{equation}
\frac{c_{o}^{[loc]}}{c_{e}^{[loc]}}=\left(\frac{a_{o}^{[loc]}}{a_{e}^{[loc]}}\right)^{-\zeta}
\end{equation}
we have:
\begin{align}
\frac{\lambda_{o}^{[loc]}}{\lambda_{o}^{*}} & =\frac{\lambda_{o}^{[loc]}}{\lambda_{e}^{[loc]}}.\frac{\lambda_{e}^{[loc]}}{\lambda_{o}^{*}}\\
 & =\frac{c_{o}^{[loc]}}{c_{e}^{[loc]}}.\frac{\nu_{e}}{\nu_{o}}.\frac{a_{e}^{[loc]}}{a_{o}^{[loc]}}\\
 & =\left(\frac{a_{o}^{[loc]}}{a_{e}^{[loc]}}\right)^{-\zeta}.\frac{a_{o}^{1+\zeta}}{a_{e}^{1+\zeta}}.\frac{a_{e}^{[loc]}}{a_{o}^{[loc]}}
\end{align}
Finally:
\begin{equation}
1+z:=\frac{\lambda_{o}^{[loc]}}{\lambda_{o}^{*}}=\left(\frac{a_{e}}{a_{o}}\right)^{-(1+\zeta)}\,\left(\frac{a_{e}^{[loc]}}{a_{o}^{[loc]}}\right)^{1+\zeta}
\end{equation}
in perfect agreement with \eqref{eq:modified-Lemaitre-1}.

\newpage{}

\end{document}